\documentclass[journal]{IEEEtran}
\usepackage{amsmath,amssymb,amsfonts}
\usepackage{graphicx}
\usepackage{breqn}
\usepackage{textcomp}
\usepackage{amsthm, bm, bbm}

\DeclareMathOperator*{\argmax}{arg\,max}
\DeclareMathOperator*{\argmin}{arg\,min}
\DeclareMathOperator\atanh{atanh}
\usepackage[font=small,labelfont=bf]{caption}
\usepackage{pdfpages, float}
\usepackage{mathrsfs}
\usepackage{graphicx}

\newtheorem{assumption}{Assumption}

\newtheorem{definition}{Definition}

\newtheorem{example}{Example}

\usepackage{subcaption}
\definecolor{blu}{RGB}{0, 102, 204}
\definecolor{purp}{RGB}{128,0,128}
\definecolor{rd}{RGB}{255,69,0}
\definecolor{org}{RGB}{255, 95, 31}
\definecolor{cyn}{RGB}{0, 200, 200}

\usepackage[linesnumbered,ruled,vlined]{algorithm2e}
\usepackage{algpseudocode}
 \usepackage{setspace}

\SetCommentSty{mycommfont}

\newcommand{\ceil}[1]{\left\lceil #1 \right\rceil}

\date{}

\begin{document}

\title{
	%\vspace{0.75cm}
	RELDEC: Reinforcement Learning-Based Decoding of Moderate Length LDPC Codes
	%\vspace{-0.35cm}
	\thanks{This work has been supported in part by U.S.~NSF grant ECCS-1711056 and the U.S.~Army Research Laboratory  under Cooperative Agreement Number W911NF-17-2-0183. This paper has been presented in part at the 17th International Symposium on Wireless Communication Systems, Berlin, Germany \cite{iswcs}.}
}

	\author{
	\IEEEauthorblockN{Salman Habib$^\dagger$, Allison Beemer$^\ast$, and J{\"o}rg Kliewer$^\dagger$\\
		}
		%\vspace{0.5cm}
	\IEEEauthorblockA{
		\begin{small}
	$^\dagger$Helen and John C.~Hartmann Dept. of Electrical and Computer Engineering, New Jersey Institute of Technology \\
	%\vspace{-0.25cm}
		$^\ast$Dept. of Mathematics, University of Wisconsin-Eau Claire %, \\Wisconsin, WI 54701
	%Email: sh383@njit.edu, dgmm@nmsu.edu, jkliewer@njit.edu
\end{small}
	}
}

\maketitle
%\vspace{-1.25cm}

\begin{abstract} 
	In this work we propose RELDEC, a novel approach for sequential decoding of moderate length low-density parity-check (LDPC) codes. The main idea behind RELDEC is that an optimized decoding policy is subsequently obtained via reinforcement learning based on a Markov decision process (MDP). In contrast to our previous work, where an agent learns to schedule only a single check node (CN) within a group (cluster) of CNs per iteration, in this work we train the agent to schedule all CNs in a cluster, and all clusters in every iteration. That is, in each learning step of RELDEC an agent learns to schedule CN clusters sequentially depending on a reward associated with the outcome of scheduling a particular cluster. We also modify the state space representation of the MDP, enabling RELDEC to be suitable for larger block length LDPC codes than those studied in our previous work. Furthermore, to address decoding under varying channel conditions, {we propose agile meta-RELDEC (AM-RELDEC)} that employ{s} meta-reinforcement learning. The proposed RELDEC scheme significantly outperforms standard flooding and random sequential decoding for a variety of LDPC codes, including codes designed for 5G new radio.	 
\end{abstract}

\vspace{-0cm}
\section{Introduction}
\label{sec:Intro}

%%%%%%%%%%%%%%%%%%%%%%%%%%%%%%%%%
%%%%%%%%%%%%%%%%%%%%%%%%%%%%%%%%%

{Low-density parity-check (LDPC) codes, a class of channel codes based on sparse parity-check matrices, are known for their excellent performance over symmetric binary input channels \cite{Gal62,Tan81, RSU01}. 
Not only can optimized LDPC codes be capacity-achieving, but as a result of their eponymous sparsity, they admit low complexity graph-based message-passing decoding algorithms, such as {belief propagation} (BP) \cite{M99}. 
Indeed, due to their performance and practical implementation, they have recently been standardized for data communication in the 5G cellular new radio standard \cite{RK18,5gnrstandard}.
It has been shown that the order in which iterative updates are sent across the code's graph edges can have significant effect on the decoder's performance \cite{YPNA01, ZF02, KK03, hocevar}. In particular, prior work has shown that sequential scheduling can reduce the number of iterations needed for convergence, hence decreasing the overall decoder complexity. 
Our own previous work in \cite{HBJNIPS, JSAIT} explored the idea of utilizing reinforcement learning (RL) in order to schedule the order of soft-information updates. 
%We found that, as compared to other sequential decoders in the literature, our learned scheduling could both improve code performance and decrease the number of iterations needed for decoder convergence on short LDPC codes. However, implementation of our work was limited by an exponentially (with block length) growing number of states, which encode the current status of the decoder's message updates. Other prior work on sequential scheduling\change{, such as \cite{CGW10},} has not utilized RL in the setting of soft updates, and has focused on designing a static scheduling order for a fixed channel.
The current work presents two main contributions: (1) we increase the feasible block length for RL-based learning of check node (CN) scheduling by introducing a new state space (i.e. the set of all possible states) for our learning algorithm and adjusting how the scheduling of CNs is executed. We refer to this scheme as RELDEC (\underline{re}inforcement \underline{l}earning-based \underline{dec}oding).
(2) We apply the improvements of RELDEC to the scenario in which channel parameters may be shifting, introducing a novel meta-learning algorithm that can adapt to new parameters with minimal additional training. We call this scheme \underline{a}gile \underline{m}eta-RELDEC, or AM-RELDEC.}

{LDPC codes may be represented by bipartite {Tanner graphs}, derived by viewing a parity-check matrix of the code as an adjacency matrix of a graph \cite{Tan81}. The vertices in one part of this graph are termed {variable nodes (VNs)}, while the vertices in the other part are {check nodes (CNs)}. Standard iterative decoding of LDPC codes is performed by passing soft probabilistic information back and forth across the edges of the Tanner graph (e.g. using the BP algorithm) via flooding: in each iteration, all CNs and VNs are updated simultaneously. In contrast, sequential {BP} decoding, also referred to as layered decoding, updates nodes individually, or as clusters, in sequence. Sequential scheduling problems are concerned with the optimal order of CN (or VN) updates, with the goal of improving the convergence speed and/or decoding performance with respect to flooding and other sequential schemes. Previous work that utilized calculations at vertices to find an optimal scheduling order includes: scheduling of CNs based on the  \textit{residual} change between iterations of the vertex's incoming messages, referred to as \textit{node-wise scheduling}, or NS \cite{CGW10}; scheduling of VNs based on a \textit{relative} residual, called \textit{efficient dynamic schedule for layered BP}, or EDS-LBP \cite{edslbp}; and VN scheduling based on a vertex's reliability as measured by incoming log-likelihood ratios (LLRs), called a \textit{reliability-based layered BP decoder}, or RBL-BP \cite{rblbp}.} 

{In our own previous work, we showed that RL can be used to improve the performance and convergence speed of sequential LDPC decoders as compared to flooding and previous node-wise scheduling schemes \cite{HBJNIPS, JSAIT}. 
%%%
Generally speaking, RL is a subset of machine learning which focuses on learning from interaction via computational means. The learning framework is comprised of a (fictitious) agent who interacts with an environment modeled by a finite Markov decision process (MDP) \cite{Sutton18}. The goal of the agent is to observe the current state of the MDP and then take an action to alter the state. After each action, the agent is rewarded based on the ``value'' of the action taken given the state. Over time, the agent learns the best action policy in order to maximize the total reward earned over time, information which is encoded using an action value function. In our work here and in \cite{HBJNIPS, JSAIT}, we utilize Q-learning, which is a Monte Carlo-based RL algorithm for learning action values \cite{Watkins89, Duff95}. Applications of RL include computer games, self-driving cars, and robotics ({e.g.}, \cite{Marina,go17,Deheng}). RL has also been applied in the context of communications (see the survey paper \cite{RLcommsurvey}). Prior applications specific to decoding include \cite{polarRL1}, which framed the factor graph selection problem of BP-based decoding of polar codes as a multi-armed bandit problem, and \cite{CHMRP19}, in which RL was used to make (hard) bit-flipping decisions in iterative decoding. The authors of \cite{CHMRP19} also suggested that RL may be applied to other decoders. Machine learning-based decoders beyond RL have also been explored: deep learning based on neural networks (NNs) was leveraged for decoding linear codes by learning the noise on the communication channel in, e.g., \cite{Nachmani_etal2018,KimetalNIPS19,Beery_etal2020}, and a deep learning framework based on hyper-networks was used for decoding short block length LDPC codes in \cite{NW19}. }

{Our work in \cite{HBJNIPS, JSAIT} differed from other RL decoders in that we used RL for \textit{soft} information BP decoding as opposed to hard-decision bit-flipping, and we sought to learn a CN update schedule as opposed to a choice of graph on which to perform decoding. More specifically, we used RL to approximate an {action value function}, which indicates the optimal next CN to schedule based on the current state of the messages being passed in decoder. In \cite{HBJNIPS}, we initiated the implementation of RL to schedule soft CN updates in an LDPC BP decoder: we considered the model-free RL methods of computing the Gittins index \cite{Git79} of each CN as well as utilizing Q-learning. In \cite{JSAIT}, we optimized the clustering of check nodes in order to improve performance, and investigated a model-based RL-NS approach with the aid of Thompson sampling. 
To the best of our knowledge, these works marked the first time that RL learning had been used to schedule sequential BP decoding with soft information updates.}

{The first main contribution of the current work is improved sequential decoding performance of longer LDPC codes using a novel RL-based scheme called RELDEC. Here, we seek to sequentially schedule {clusters} of CNs: in a Tanner graph containing $m$ CNs with each cluster comprised of $z$ CNs, the scheduling problem is modeled as a finite MDP with $\ceil{m/z}$ possible actions (i.e. cluster selections). We note that the cluster size $z$ should be selected appropriately to balance the trade-off between decoding performance and throughput. As we will see later, smaller clusters will lead to lower Q-learning complexity and improved performance. Thus, we choose cluster sizes no larger than $2$ in our experiments. RELDEC learns an action value function that determines how beneficial a particular choice of cluster is for optimizing the overall cluster scheduling policy. The learned policy is then incorporated into our sequential LDPC decoding algorithm for inference. 
A limitation of our previous work was that we were only able to implement our scheduling strategy up to block length 196. This is due to the fact that for MDPs with a large state space, Q-learning can require tremendous computational effort. 
A multitude of methods for reducing this type of learning complexity have been proposed: for example, partitioning the state space ({e.g.,} \cite{csimcsek2005identifying,mannor2004dynamic}), imposing a state hierarchy ({e.g.,} \cite{parr1998reinforcement}), or {reducing dimensionality} ({e.g.,} \cite{BitzerHowardVijayakumar2010, ShahXie2018}). One of the salient features of RELDEC is a significant reduction in the cardinality of the state space due to a new method of identifying the states of our clusters as compared with our previous RL-NS scheme; this enables the decoding of LDPC codes with block lengths of up to 500 bits. Another distinct feature of our current work is that every cluster is scheduled in each executed iteration. In our previous work, a CN cluster was scheduled in each iteration, without any limitation on the possibility of one cluster being scheduled multiple times before another cluster is scheduled for the first time. This change increases the exploration of the RL agent, resulting in a lower BER in the error floor region.  Simulations reveal that RELDEC results in significant decoding gain for fixed bit error rate (BER) compared to traditional BP decoding schemes such as flooding, NS, EDS-LBP, RBL-BP, and a variable-node-based layered decoding scheme proposed in the 5G standardization process in \cite{5gdoc}. Moreover, RL-based decoders require a smaller number of CN to VN message updates, on average, to achieve these gains.}

{Leveraging our updated state space and scheduling policy, we turn to our second main contribution, where we investigate the scenario in which the parameters of a channel may shift over space and/or time. Prior scheduling policies (including our own) have focused on determining an optimal fixed scheduling order for one particular channel. However, a fixed update order, even one that is optimized for an initially-observed channel, may not provide the optimal CN scheduling policy when a new signal-to-noise ratio (SNR) is encountered by the decoder. To address such a setting, we propose a novel Q-learning-based meta-learning scheme called agile meta-RELDEC, or AM-RELDEC. }

{Meta-learning has gained considerable attention in recent years. At a high level, meta-learning algorithms first accumulate experience in solving a variety of tasks, and then adapt for solving related but unseen tasks. During adaptation, the algorithm is expected to perform well on the new tasks given minimal training steps and data. A model-agnostic meta-learning (MAML) algorithm was proposed in \cite{Finn} for a wide range of problems including classification, regression, and RL. The MAML scheme relies on gradient descent for model parameter optimization. In \cite{Fakoor}, the authors proposed a meta-Q-learning (MQL) scheme, a gradient-based approach suitable for Q-learning-based meta-learning. Meta-learning has also been used to solve problems related to communication systems: for instance, MAML was proposed for designing an optimized demodulator which quickly adapts to various channel conditions after being trained using a small number of pilot symbols in \cite{Simeone,Park}. This demodulation scheme is suitable for an internet-of-things (IoT) setting, where the channel conditions may vary between devices. MAML was also implemented in the context of channel coding in \cite{Li21}: the work's supervised meta-learning scheme learns to map a received noisy signal to an estimation of the transmitted message. While AM-RELDEC, which is described in more detail below, also addresses a decoding problem, we note that the block length of 20 considered in \cite{Li21} is substantially smaller than the block lengths considered in this work.}

{In our AM-RELDEC scheme, we learn a global action value function corresponding to CN scheduling using data from a mixture of SNRs, then update this action value function locally given an observed (instantaneous) SNR. Once the global action value function has been optimized, adaptation to a local SNR is achieved in an online fashion, based on a minimal number of additional training vectors. In contrast to RELDEC, the agility of AM-RELDEC allows for dynamic scheduling which can adapt to unseen SNRs with substantially reduced additional training using, e.g., pilot signals. Due to its flexibility, AM-RELDEC is well-suited for a multitude of IoT applications. As an example, consider the setting of smart transportation \cite{smartv}. In this scenario, the power of a signal received by a vehicle from a road side unit (RSU) changes as the vehicle moves, causing fluctuations in channel SNR.  During communication with an RSU, the vehicle's sensor can detect the instantaneous SNR of the channel and, using AM-RELDEC, would be able to quickly adapt the decoder's CN scheduling policy. In addition to this nimbleness, AM-RELDEC inherits the gains displayed by RELDEC.}

{The paper is organized as follows: background on LDPC codes and reinforcement learning is is given in Section \ref{sec:prelim}. In Section \ref{sec:RL}, we describe how our CN scheduling policy is learned via RELDEC and then incorporated into our learning-based sequential decoding algorithm. Section \ref{sec:meta-RL} {discusses how our scheduling policy is learned using AM-RELDEC}. In Section \ref{sec:results}, we explain our experimental setup and analyze numerical results, comparing the proposed learning-based decoding schemes to LDPC decoders found in the literature. Section \ref{sec:concl} concludes the paper.}

%%%%%%%%%%%%%%%%%%%%%%%%%%%%%%%%%
%%%%%%%%%%%%%%%%%%%%%%%%%%%%%%%%%

\vspace{-0.cm}
\section{Preliminaries}
\label{sec:prelim}

\subsection{Low-density Parity-check Codes}

An {$[n,k]$ binary linear code} is a $k$-dimensional subspace of $\mathbb{F}_{2}^{n}$, and may be defined as the kernel of a binary \emph{parity-check matrix} $\mathbf{H}\in \mathbb{F}_2^{m\times n}$, where $m\geq n-k$. The code's \emph{block length} is  $n$, and \textit{rate} is $(n-\text{rank}(\mathbf{H}))/n$.  The \emph{Tanner graph} of a linear code with parity-check matrix $\mathbf{H}$ is the bipartite graph $G_{\mathbf{H}}=(V\cup C,E)$, where $V=\{v_0,\ldots,v_{n-1}\}$ is a set of VNs corresponding to the columns of $\mathbf{H}$, $C=\{c_0,\ldots,c_{m-1}\}$ is a set of CNs corresponding to the rows of $\mathbf{H}$, and edges in $E$ correspond to $1$'s in $\mathbf{H}$ \cite{Tan81}. LDPC codes are a class of highly competitive linear codes defined via sparse parity-check matrices or, equivalently, sparse Tanner graphs \cite{Gal62}; they are amenable to low-complexity graph-based message-passing decoding algorithms, making them ideal for practical applications. BP iterative decoding, considered here, is one such algorithm.

%In this work, we present experimental results for a standardized $[384,256]$-Wireless Regional Area Network (WRAN) LDPC code \cite{wran} and a
%two particular classes of LDPC codes: $(\gamma,k)$-regular 
%$(\gamma,p)$-regular array-based (AB-) LDPC code. \ab{It's not clear why these two are stated here and 5G-NR is stated later, especially since we don't talk about the WRAN construction. Maybe introduce all three codes used at the end of the previous paragraph, then give some details on each in subsequent paragraphs?}

In general, a $(\gamma,k)$-regular LDPC quasi-cyclic (QC) code is defined by a parity-check matrix with constant column and row weights equal to $\gamma$ and $k$, respectively \cite{fossqc}. A $(\gamma,p)$ array-based (AB) LDPC code is a type of QC code where $p$ is prime, and its parity-check matrix possess a special structure \cite{Fan00}. In particular, for AB codes, 

\begin{equation}
\label{eq:mat}
\mathbf{H}(\gamma,p)=
\begin{bmatrix}
\mathbf{I} & \mathbf{I} & \mathbf{I} & \cdots & \mathbf{I} \\
\mathbf{I} & \sigma & \sigma^{2} & \cdots & \sigma^{p-1} \\
\vdots & \vdots & \vdots & \cdots & \vdots \\
\mathbf{I} & \sigma^{\gamma-1} & \sigma^{2(\gamma-1)} & \cdots & \sigma^{(\gamma-1)(p-1)}
\end{bmatrix},
\end{equation}

%\vspace{0.5cm}
\noindent where $\sigma^z$ denotes the circulant matrix obtained by cyclically left-shifting the entries of the $p\times p$ identity matrix $\mathbf{I}$ by $z$ (mod $p$) positions. Notice that $\sigma^0=\mathbf{I}$. In this work, \textit{lifted} LDPC codes are obtained by replacing the non-zero (resp., zero) entries of the parity-check matrix {with} randomly generated permutation (resp., all-zero) matrices.

%Each row (resp.,~column) of sub-matrices of $\mathbf{H}(\gamma,p)$ forms a \emph{row} (resp.,~\emph{column}) \emph{group}. Observe that there are a total of $p$ (resp.,~$p^2$) column groups (resp.,~columns) and $\gamma$ (resp.,~$\gamma p$) row groups (resp.,~rows) in $\mathbf{H}(\gamma,p)$. 

Parity-check matrices of LDPC codes designed for 5G new radio (5G-NR) possess a QC structure which enables the nodes in its Tanner graph to be updated in parallel \cite{5g}. According to the third generation partnership project (3GPP) standard, 5G-NR LDPC codes can be obtained by lifting two types of QC base graphs, known as BG1 and BG2. Depending on the lifting factor, BG1 can be used for constructing LDPC codes with information bits ranging between $500$ and $8448$ bits, whereas the BG2 base matrix can be used for obtaining shorter codes with information bits ranging between $40$ and $2560$ bits \cite{5g,Bae}. The BG1 (resp., BG2) matrix is used for rates between $1/3$ and $8/9$ (resp., $1/5$ and $2/3$) \cite{5g,Bae}. 

%Since the amount of resources available for transmission can change dynamically in a cellular system, 5G-NR LDPC codes are required to have rate matching functionality to select an arbitrary amount of information bits for transmission. The rate of a 5G-NR LDPC code can be increased by puncturing CNs in the base graph. 

%Each base matrix consists of five sub-matrices $A$, $B$, $C$, $D$ and an identity matrix $E$. The columns of $A$ correspond to the systematic portion of the matrix, and $B$ is a square matrix with a dual-diagonal structure. Sub-matrix $C$ is an all-zero matrix whose columns correspond to the extended VNs of the code. The $A$, $B$ and $C$ sub-matrices share the same rows and they correspond to the core CNs of the code. Sub-matrices $D$ and $E$ share the same rows and they correspond to the extended CNs of the code.  The structure of a 5G-NR LDPC base parity-check matrix is shown in Fig. ...

%The systematic information bits corresponding to the first two columns of the base matrix are punctured and are not transmitted.

\subsection{Reinforcement Learning}
%\vspace{-0.1cm}

In an RL problem, an agent (learner) interacts with an environment whose \textit{state space} can be modeled as a finite MDP \cite{Sutton18}. The agent takes \textit{actions} that {alter} the state of the environment and receives a \textit{reward} in return for each action, with the goal of maximizing the total reward in a series of actions. The optimized sequence of actions is obtained by employing a policy which utilizes an \textit{action value function} to determine how beneficial an action is for maximizing the long-term expected reward. In the remainder of the paper, let $[[x]]\triangleq \{0,\ldots,x-1\}$, where $x$ is a positive integer. Suppose that an environment allows $m$ possible actions, and let {the random variable} $A_\ell\in [[m]]$ with realization $a$ represent the index of an action taken by the agent during learning step {$\ell\in \{0,\ldots, \ell_\max-1\}$}. Let $S_{\ell}$ with realization $s^{(\ell)}\in \mathbb{Z}$ represent the current state of the environment before taking action $A_\ell$ and let $S_{\ell+1}$ with realization $s^{(\ell)'}$ represent a new state of the MDP after executing $A_\ell$. Let a state space $\mathcal{S}$ contain all possible state realizations. Also, let $R_\ell=R(S_{\ell},A_\ell,S_{\ell+1})$ be the reward yielded at step $\ell$ after taking action $A_\ell$ in state $S_{\ell}$ which will yield state $S_{\ell+1}$. {Optimal policies for MDPs can be estimated via Monte Carlo techniques, such as Q-learning. The estimated action value function $Q_\ell(S_{\ell},A_\ell)$ (Q-function) represents the expected long-term reward an agent obtains after taking action $A_\ell$ in state $S_{\ell}$. Learning involves iteratively adjusting the action value function for a specific  $(S_{\ell},A_\ell)$ pair, based on  previously learned action values for the same state and action pair and the reward $R_\ell$ earned from taking that action. The optimal policy guides the agent to select an action in a given state that maximizes the Q-function value for that state.}

%(see (\ref{eq:q_cls2})), (see (\ref{eq:pol_clus}))

%In meta-learning, bias refers to a set of assumptions which influences the selection of hypothesis for a given learning task. Bias can be classified as a declarative bias, which specifies the representation of the space of hypothesis (for instance, representing hypothesis using artificial NNs), or procedural bias, which is concerned with the ordering or hypothesis (for instance, preferring hypothesis with a smaller run-time) . 

%\section{RL for Sequential LDPC Decoding}
%\label{sec:RLSD}
\section{Learning the Scheduling Policy Using RELDEC}
\label{sec:RL}

The proposed RELDEC scheme consists of a BP decoding algorithm in which the environment is given by the Tanner graph of the LDPC code. {Our objective is to learn an  optimized sequence of actions, \emph{i.e.}, the scheduling of individual sets $\mathcal{C}_1,\ldots,\mathcal{C}_{\ceil{m/z}}$ of CNs with the  set of CNs $\mathcal{C}_i=\{c_{i,1},\ldots,c_{i,z}\}$ of cardinality $z=|\mathcal{C}_i|$}. Herein,  $c_{i,j}\in [[m]]$, the index of the $j$-th CN in the $i$-th set, is called a \emph{cluster} in the remainder of the paper. By $\mathcal{N}(\mathcal{C}_i)$ we refer to all the VNs connected to the $i$-th cluster. 
%The optimized cluster scheduling order is learned via the %Q-learning scheme}. 
A single cluster scheduling step is carried out by sending messages from all CNs of a cluster to their neighboring VNs, and subsequently sending messages from these VNs to their CN neighbors. That is, a selected cluster executes one iteration of localized flooding in each decoding instant. Every cluster is scheduled exactly once within a single decoder iteration. Sequential cluster scheduling is carried out until a stopping condition is reached, or an iteration threshold is exceeded. The learning-based decoder relies on a cluster scheduling policy based on a learned action value function. %, which is estimated offline using the RL techniques to be discussed in further detail in Section \ref{sec:RL}. 

%The RELDEC learning framework for sequential decoding is shown in Fig. \ref{fig:RL}. The idea %is that scheduling a cluster, represented by the blue squares (CNs), updates the state of the %environment, namely the hard-decisioned beliefs associated to the blue VNs connected to the %cluster. In return, the agent receives a reward which is commensurate with the proportion of %correct hard decisions. %The accumulation of rewards over time allows the agent to optimize %the cluster scheduling order.

\begin{figure}[h]
  \centering
  \includegraphics[scale=0.48]{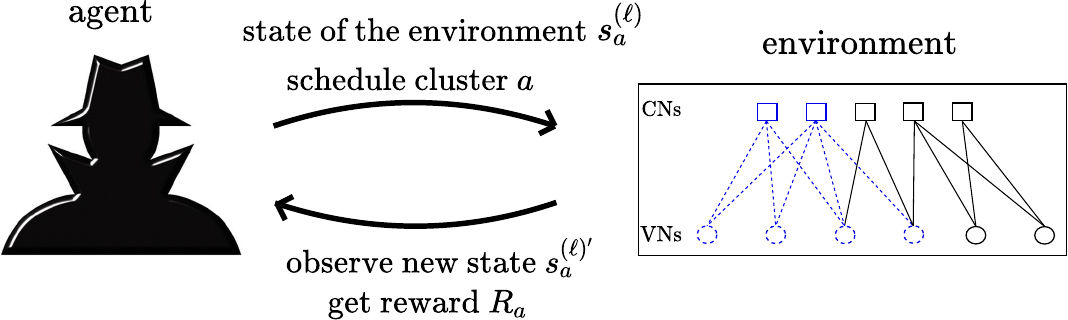}
  %\vspace{-0.5cm}
  \caption{Illustration of RELDEC's learning framework. In each learning step, a fictitious agent schedules a cluster with index $a$ when the environment state, based on hard-decisioned VN values, is $s$. Once an action is taken, the state of the environment changes from $s_a^{(\ell)}$ to $s_a^{(\ell)'}$ as the VN values are updated after scheduling, and the agent receives reward $R_a$ that indicates the accuracy of the hard-decisions taken by the BP algorithm for each blue VN.}
  \label{fig:RL}
\end{figure}

{The MDP related to the RELDEC learning framework for sequential decoding is shown in Fig. \ref{fig:RL}}. The idea is that scheduling a cluster, represented by the blue CNs, updates the state of the environment. In return, the agent receives a reward which is commensurate with the proportion of correct hard decisions. Here, the state is determined by the union of two quantities: the set of check-node indices belonging to a cluster and a binary vector given by the hard-decisioned beliefs of the VNs associated with the cluster.

%\vspace{-0.5cm}

\begin{figure}[h]
  \centering
  %\centerline{\resizebox{3in}{2.4in}{\includegraphics{pics/res1d.pdf}}}
  \includegraphics[scale=1.1]{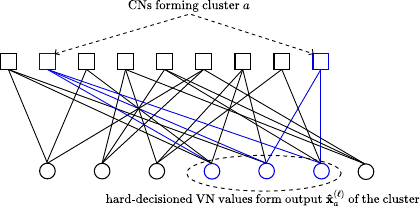}
  %\vspace{-0.12cm}
  \caption{Example of a cluster-induced subgraph, shown with blue squares (cluster consisting of $2$ CNs), edges, and circles (VNs). The corresponding state of the cluster is $\mathbf{\hat{x}}_a^{(\ell)}$.} 
  \label{fig:cls}
\end{figure}

{We define the action space of the MDP as $\mathcal{A}=[[\ceil{m/z}]]$; this set has a cardinality of $\ceil{m/z}$. For example, for $m=5$ and $z=2$, $\mathcal{A}=\{0,1,2\}$.} Let $\mathbf{\hat{x}}_a^{(\ell)}= [\hat{x}_{0,a}^{(\ell)},\ldots,\hat{x}_{l_a-1,a}^{(\ell)}]{ \in \{0,1\}^{l_{a}}}$ denote the state of the MDP after scheduling a cluster with index $a\in \mathcal{A}$ during learning step $\ell$, and let $s_a^{(\ell)}\in [[2^{l_a}]]$ refer to the index of a realization of $\mathbf{\hat{x}}_a^{(\ell)}$. Thus, $s_a^{(\ell)}$ also refers to the state of the MDP during learning step $\ell${, and the state space of a cluster is the set of all possible values of $s_a^{(\ell)}$. At a particular decoder iteration, let the \textit{output} of a cluster  be the binary sequence resulting from hard-decisions on the posterior LLRs computed by the (ordered) neighboring VNs.} Since the state space of the clusters are pairwise disjoint, {the overall state space} $\mathcal{S}$ of our MDP contains $\sum_{a\in [[\lceil m/z\rceil ]]}2^{l_{a}}$ realizations of all the cluster outputs $\mathbf{\hat{x}}_0^{(\ell)},\ldots,\mathbf{\hat{x}}_{\ceil{m/z}-1}^{(\ell)}$, where a realization can be thought of as a (cluster, cluster state) pair. {Note that the cluster size $z$ should be selected appropriately to balance the trade-off between decoding performance and throughput. For a cluster size of $z=1$ we schedule only one CN, and, as we show later in Section~V, achieve the best performance. In contrast, $z=m$ leads to a flooding schedule which has the highest throughput, but only a modest performance. In this paper, we focus on improving the performance and choose cluster sizes no larger than $2$ in our experiments. To facilitate a general treatment, we use an arbitrary $z\in[[m]]$ in the following.}  An example of a cluster-induced subgraph for the case $z=2$, and the corresponding state vector $\mathbf{\hat{x}}_a^{(\ell)}$ is shown in Fig. \ref{fig:cls}.

%Note that the state space size of the MDP, and hence the complexity of the RELDEC and meta-RELDEC schemes, can grow exponentially with the number of CNs, which can range in the hundreds for practical LDPC codes. 

%The state of the MDP in both our RELDEC and meta-RELDEC schemes are then given by the collection of all possible (cluster, cluster state) pairs.

%Note that in the absence of clustering (\emph{i.e.,} $z=m$), the state space cardinality would be $2^n$, which for moderate length codes studied in this paper is prohibitively large. {Creating clusters of CNs significantly reduces the state space cardinality of the MDP, which reduces the learning complexity in turn.}

Let $\mathbf{x}=[x_0,\ldots,x_{n-1}]$ and $\mathbf{y}=[y_0,\ldots,y_{n-1}]$ represent the transmitted and the received words, respectively, where for each $v\in [[n]]$, $x_v\in \{0,1\}$ and $y_v=(-1)^{x_v}+z$ with $z\sim \mathcal{N}(0,\sigma^2)$. The posterior LLR of $x_v$ is expressed as $L_v=\log \frac{\Pr(x_v=0|y_v)}{\Pr(x_v=1|y_v)}$. Let $\hat{L}_\ell^{(v)}=\sum_{c\in \mathcal{N}(v)} m_{c\rightarrow v}^{(\ell)}+L_v$ be the posterior LLR computed by VN $v$ during iteration $\ell$, where $\mathcal{N}(v)$ denotes the set of neighboring CNs of VN $v$, $\hat{L}_0^{(v)}=L_v$,  and $m_{c\rightarrow v}^{(\ell)}$ is the message received by VN $v$ from neighboring CN $c$ in iteration $\ell$ computed based on standard BP as 

\begin{equation}
	m_{c\rightarrow v}^{(\ell)}=2\atanh \prod_{v'\in \mathcal{N}(c)\setminus v} \tanh\left(\frac{m_{v'\rightarrow c}^{(\ell-1)}}{2}\right).
	\label{eq:mcv}
\end{equation}
Here, $\mathcal{N}(c)$ denotes the set of neighboring VNs of $c$, and
\vspace{-0.25cm}
\begin{equation}
	m_{v\rightarrow c}^{(\ell)}= \sum_{c'\in \mathcal{N}(v)\setminus c} m_{c'\rightarrow v}^{(\ell)}+L_v
	\label{eq:mvc}
\end{equation}
is the message propagated from VN $v$ to CN $c$. Moreover, let $\hat{L}_\ell^{(j,a)}$ be the posterior LLR computed during learning step $\ell$ by VN $j$ in the subgraph induced by the cluster with index $a\in[[\ceil{m/z}]]$. Hence, $\hat{L}_\ell^{(j,a)}=\hat{L}_\ell^{(v)}$ if VN $v$ in the Tanner graph is also the $j$-th VN in the subgraph induced by the cluster with index $a$. 

Note  that in our prior work \cite{JSAIT}, the RL-NS algorithm scheduled a single CN $a$ per decoding iteration based on its reward $\max_{v\in \mathcal{N}(a)}$ $r_{a\rightarrow v}$, where $r_{a\rightarrow v}$ is the residual of CN $a$ associated with the edge connecting to VN $v$, computed according to $r_{a\rightarrow v}\triangleq|m_{a\rightarrow v}'-m_{a\rightarrow v}|$. Here, $m_{a\rightarrow v}$ is the message sent by CN $a$ to its neighboring VN $v$ in the previous iteration, and $m_{a\rightarrow v}'$ is the message that CN $a$ would send to VN $v$ in the current iteration, if scheduled. {So, the residual represents a quantity which is associated with each edge of the Tanner graph for each BP iteration.} Intuitively, the higher the residual of a CN, the further away that portion of the graph is from convergence. Thus, scheduling a CN with the highest residual (reward) leads to faster and more reliable decoding compared to the flooding scheme. Furthermore, in \cite{JSAIT}, the state space of the MDP is given by the collection of all sequences representing quantized CN values within a cluster.
%$\mathcal{N}(a)$ is the set of all neighboring VNs of $a$, and 

In contrast, in the proposed RELDEC scheme we consider a new state space representation of the MDP {along with a different reward computation}, which allows learned decoding of significantly longer block-length LDPC codes. In RELDEC, after scheduling cluster $a$ during learning step $\ell$, the state of the MDP associated with cluster $a$ is given by its output $\mathbf{\hat{x}}_a^{(\ell)}$ that is obtained by taking hard decisions on the vector of posterior LLRs $\mathbf{\hat{L}}_{\ell,a}=[\hat{L}_\ell^{(0,a)} \ldots \hat{L}_\ell^{(l_a-1,a)}]$, computed according to 

\vspace{-0.25cm}
\begin{equation}
	\hat{x}_{j,a}^{(\ell)}=
	\begin{cases}
		0, \text{ if } \hat{L}_\ell^{(j,a)} \geq 0,\\
		1, \text{ otherwise,}
	\end{cases}
	\label{eq:hd}
\end{equation}
{where $k_\max$  is the maximum CN degree of the cluster, and $l_a\leq k_\max z$ is the number of VNs adjacent to cluster $a$. }
%where $l_a\leq k_\max z$ is the maximum CN degree of the cluster, and $k_\max$ is the number of VNs adjacent to cluster $a$. 
We call $\mathbf{\hat{x}}_a^{(\ell)}$ the state of cluster $a$: it is comprised of the bits reconstructed by the sequential decoder after scheduling cluster $a$ during iteration $\ell$, {\emph{i.e.,} the state of the cluster is a sequence of hard-decision VN values associated with the cluster.} The collection of signals $\mathbf{\hat{x}}_0^{(\ell)},\ldots,\mathbf{\hat{x}}_{\ceil{m/z}-1}^{(\ell)}$ at the end of decoder iteration $\ell$ forms the entire state of the MDP associated with our RELDEC scheme. 

%We denote the index of a realization of $\mathbf{\hat{x}}_a^{(\ell)}$ in iteration $\ell$ by $s_a^{(\ell)}\in [[2^{l_{a}}]]$.
 
During the learning phase, RELDEC informs the agent of the current state of the decoder and the reward obtained after performing an action (propagating messages from a cluster to its neighboring VNs). Based on these observations, the agent takes future actions, to enhance the total reward earned, which alters the state of the environment as well as the future reward. Given that the transmitted signal $\mathbf{x}$ is known during the training phase, let $\mathbf{x}_a=[x_{0,a},\ldots,x_{l_a-1,a}]$ be a vector containing the $l_a$ bits of $\mathbf{x}$ that are reconstructed in $\mathbf{\hat{x}}_a^{(\ell)}$ by cluster $a$. {Note that  is the corresponding state vector after scheduling the check nodes in $a$.} In each learning step $\ell$, the reward $R_a$ obtained by the agent after scheduling cluster $a$ is defined as 

\vspace{-0.25cm}

\begin{equation}
	\label{eq:rew}
	R_a=\frac{1}{l_a} \sum_{j=0}^{l_a-1} \mathbbm{1}(x_{j,a} = \hat{x}_{j,a}),
	%R_a=1-\Pr[x_{j,a}\neq \hat{x}_{j,a}].
\end{equation}
where $\mathbbm{1(}\cdot)$ denotes the indicator function. Thus, the reward earned by the agent after scheduling cluster $a$ is identical to the probability that the transmitted bits $x_{0,a},\ldots,x_{l_a-1,a}$ are correctly reconstructed. This new reward metric, resulting from the modification of the state space representation, differs considerably from the maximum residual of the scheduled CN used as reward for RL-NS in \cite{JSAIT}. 

The action values learned by RELDEC are stored in a table with dimension $\max_a(2^{l_a})\times \ceil{m/z}$. In comparison, for the RL-NS scheme of \cite{JSAIT}, the learned action values are stored in a table with dimension $M^{z}\times \ceil{m/z}$, where $M=4$ is the number of quantization levels used for quantizing the CN values, and $M^z$ is the number of all possible sequences of quantized CN values associated with a cluster. Thus, for the MDP discussed in our previous work the state space cardinality, and hence the learning complexity, grows exponentially with $z$. Furthermore, a moderately large $z$ ($\geq 7$) was chosen to ensure a small number of clusters, since there exist dependencies between clusters (\emph{i.e.,} the state of one cluster may depend on the state of another cluster due to the presence of cycles) which Q-learning cannot take into account. However, due to the modification of the MDP in this work, even a choice of $z=1$ provides considerable reduction of the state space, and hence the  size of the action value table is also significantly reduced. This yields a reduced learning complexity with respect to the RL-NS scheme in \cite{JSAIT}. 

%\sum_{a\in [[\lceil m/z\rceil ]]}, \ceil{m/z}

In the following, we discuss the learning approach used by RELDEC for obtaining the optimal CN scheduling policy for a given LDPC code. As the new state space representation generates MDPs with moderately large state space size, we utilize standard Q-learning for determining the optimal cluster scheduling order, where the action value, $Q_{\ell+1}(s_a^{(\ell)},a)$, for choosing cluster $a$ in state $s_a$ is given by
\begin{align}
	\begin{split}
	Q_{\ell+1}(s_a^{(\ell)},a)= (1-\alpha)&Q_\ell(s_a^{(\ell)},a) + \\
	&\alpha \Bigl(R_a+\beta \max_{a'\in [[\ceil{m/z}]]}Q_\ell(s_a^{(\ell)'},a') \Bigr),
	\label{eq:q_cls2}
	\end{split}
\end{align}
{\noindent where $s_a^{(\ell)'}$ represents the new state of the MDP after taking action $a$ in state $s_a^{(\ell)}$, $0<\alpha <1$ is the \textit{learning rate}, $0<\beta<1$ is the \textit{reward discount rate}, $Q_{\ell+1}(s_a^{(\ell)},a)$ is a future action value resulting from action $a$ in the current state $s_a^{(\ell)}$ \cite{Watkins89}, and $\ell$ is the number of learning steps elapsed after observing the initial state $s_a^{(0)}$, corresponding to a received channel output $\mathbf{L}=[L_0,\ldots,L_{n-1}]$, in a learning episode\footnote{A learning episode comprises all the Q-learning steps needed to learn the action values corresponding to a single training example.}. For each $\ell$, cluster $a$  is selected via an $\epsilon$-greedy approach according to} 
 \begin{equation}
	\label{eq:egreedy}
	a=
	\begin{cases}
     \text{selected uniformly at random w.p. } \epsilon \text{ from }\mathcal{A}, \\
	 \pi(s_a^{(\ell)}) \text{ selected w.p. } 1-\epsilon, 
	\end{cases}
\end{equation}
{\noindent where $\epsilon$ is the probability of exploration, $\mathcal{A}=\{0,\ldots,\ceil{m/z}\}$ is a set of all possible actions, and $\pi(s_a^{(\ell)})$ is an agent's  policy for taking an action in state $s_a^{(\ell)}$ expressed as}
\begin{equation}
	\label{eq:pol_clus}
	\pi(s_a^{(\ell)})=\argmax_{a \in [[\ceil{m/z}]]} Q_\ell(s_a^{(\ell)},a).
\end{equation}
{Note that $\epsilon$ should be large enough to allow adequate exploration, but not so large which inhibits exploitation, \emph{i.e.,} taking actions according to $\pi(s_a^{(\ell)})$. Hence, $\epsilon$ should be chosen carefully to balance this trade-off. The action value function is recursively updated $\ell_{\max}$ times according to (\ref{eq:q_cls2}) after observing the initial state. The goal of Q-learning is to find the optimal policy that maximizes the long-term expected reward in state $s_a^{(\ell)}$, given by 
% \vspace{-0.25cm}
\begin{equation}
	\label{eq:pi_Qopt}
	\pi^*(s_a^{(\ell)})=\argmax_{a} Q^*(s_a^{(\ell)},a),
\end{equation}
% \vspace{-0.25cm}
 where $Q^*(s_a^{(\ell)},a)$ is the optimal action value for a given $(s_a^{(\ell)},a)$ pair.
}

For ties (as in the first iteration of Algorithm \ref{alg:Qlrn} for $\ell=0$ and the initial $\mathbf{L}$), we choose an action uniformly at random from all the maximizing actions. During inference, the optimized cluster scheduling policy of standard Q-learning, $\hat{\pi}(s_{a_i}^{(I)})$, for scheduling the $i$-th cluster during decoder iteration $I$ is expressed as 
\begin{equation}
	\label{eq:pi_i2}
 \hat{\pi}(s_{a_i}^{(I)})=\argmax_{a_i\in [[\ceil{m/z}]]\setminus \{a_0,\ldots, a_{i-1}\}} \hat{Q}(s_{a_i}^{(I)},a_i), 
\end{equation} 
where $i\in [[\ceil{m/z}]]$, and $a_i$ indicates the cluster index to be scheduled at time instant $i$. Further, $\hat{Q}(s_{a_i}^{(I)},{a_i})$ represents the optimized action value after training has been accomplished, which, as $\ell \rightarrow \infty$, {approaches} the optimal action value $Q^*(s_{a_i}^{(I)},{a_i})$ \cite{sutton88}, \cite[Sec. 6.4]{Sutton18}. The RELDEC scheme, which employs standard Q-learning, is shown in Algorithm \ref{alg:Qlrn}. The input to this algorithm is a parity-check matrix $\mathbf{H}$ and a set $\mathscr{\hat{L}}=\{\mathbf{L}_0,\ldots,\mathbf{L}_{\mathscr{|\hat{L}}|-1}\}$ containing $|\mathscr{\hat{L}}|$ realizations of $\mathbf{L}$ over which Q-learning is performed. {Note that each vector in $\mathscr{\hat{L}}$ correspond to an SNR selected from a set $\mathcal{S}=\{S_1,\ldots,S_K\}$, where $S_i\in \mathbb{R}$ is the $i$-th SNR value, and $K$ is the total number of distinct SNRs considered for training. There are $|\mathscr{\hat{L}}|/K$ LLR vectors in $\mathscr{\hat{L}}$ with the same SNR. For a $\mathbf{L} \in {|\mathscr{\hat{L}}|}$, the action value function in (\ref{eq:q_cls2}) is recursively updated $\ell_{\max}$ times as shown in Step 20.} The output of RELDEC is an optimized cluster scheduling policy $\hat{\pi}(s_{a_i}^{(I)})$.

{\linespread{0.85}\selectfont  
\begin{algorithm}
%\small
\caption{RELDEC}
\SetAlgoLined
\DontPrintSemicolon
\SetKwInOut{Input}{Input}
\SetKwInOut{Output}{Output}
%\underline{function} $\text{IPA}(\boldsymbol{y},\mathbf{A})$\;
\Input{set of channel information vectors $\mathscr{\hat{L}}$, parity-check matrix $\mathbf{H}$}
\Output{optimized cluster scheduling policy $\hat{\pi}(s_{a_i}^{(I)})$} %for $u\in 0,\ldots, \ceil{\frac{m}{z}}-1$, $s_u\in 0,\ldots, M^z-1$, and $a_u\in0,\ldots,z-1$\;
\label{alg:Qlrn}

Initialization: $Q_0(s_a^{(0)},a)\leftarrow 0$ for all $s_a^{(0)}$ and $a$
\hspace{0.25cm} 

%\tcp{clustered Q-learning starts}
\For{each $\mathbf{L}\in \mathscr{\hat{L}}$} { 
	$\ell \leftarrow 0$\;
	{$\mathbf{\hat{L}}_\ell\leftarrow \mathbf{L}$}\;
	{determine initial states of all clusters using (\ref{eq:hd})}\;
	\tcp{start of an episode}
	\While{$\ell<\ell_{\max}$}
	{
	select cluster $a$ according to {(\ref{eq:egreedy})}\;
	%{decode cluster induced subgraph via flooding}\;
	{for each CN $c$ in $\mathcal{C}_a$, compute and propagate $m_{c\rightarrow v}^{(\ell)}$ $\forall v\in \mathcal{N}(c)$\;}
	{for each VN $v$ in $\mathcal{N}(\mathcal{C}_a)$, compute and propagate $m_{v\rightarrow c}^{(\ell)}$ $\forall c\in \mathcal{N}(v)$\;}
%	\tcp{decode cluster via flooding}
%	\ForEach{CN $c$ in cluster $a$} {
%		\ForEach{VN $v\in \mathcal{N}(c)$}{
%				
%			compute and propagate $m_{c\rightarrow v}^{(\ell)}$\;
%		}
%	}
%		
%	\ForEach{VN $v$ in the subgraph of cluster $a$} {
%		\ForEach{CN $c\in \mathcal{N}(v)$}{
%				
%			compute and propagate $m_{v\rightarrow c}^{(\ell)}$\;
%		}
%
%	$\hat{L}_\ell^{(v)}\leftarrow \sum_{c\in \mathcal{N}(v)} m_{c\rightarrow v}^{(\ell)}+L_v$ \tcp*{update posterior LLR}
%}
	\tcp{determine cluster output}
		\ForEach{VN $v$ in the subgraph of cluster $a$}{
			\If{$\hat{L}_\ell^{(v)}\geq 0$}{ 
				$\hat{x}_{v,a}^{(\ell)}\leftarrow 0$\; %tcp*{i.e. use (\ref{eq:pi_G}), (\ref{eq:pi_Q})}
	}
\Else{$\hat{x}_{v,a}^{(\ell)}\leftarrow 1$\;
}
}
		determine index $s_a^{(\ell)'}$ of $\mathbf{\hat{x}}_a$\;
		update $R_a$ according to (\ref{eq:rew})\;
		compute $Q_{\ell+1}(s_a^{(\ell)},a)$ according to (\ref{eq:q_cls2})\;
		$s_a^{(\ell+1)}\leftarrow s_a^{(\ell)'}$\;
		$\ell\leftarrow \ell+1$\;
	}
	%}
}
\end{algorithm}
}

%Once the optimized scheduling policy, $\hat{\pi}(s_{a_i}^{(I)})$, is learned using Algorithm \ref{alg:Qlrn}, it is incorporated in our RL-based sequential BP decoding scheme shown in Algorithm \ref{alg:RL-SD} to determine the optimized cluster scheduling order. 

{Once learning ends, we utilize Algorithm \ref{alg:RL-SD} for inference. The algorithm inputs are the soft channel information vector $\mathbf{L}$, that corresponds to one of the SNRs in $\mathcal{S}$, and a parity-check matrix $\mathbf{H}$ of the LDPC code, and $\mathbf{\hat{L}}_{I}=[\hat{L}_I^{(0)},\ldots,\hat{L}_I^{(n-1)}]$ is initialized using $\mathbf{L}$.} The optimized scheduling policy, $\hat{\pi}(s_{a_i}^{(I)})$, is selected in Step 9 of Algorithm~\ref{alg:RL-SD}  according to (\ref{eq:pi_i2}); \emph{i.e.,} an optimized cluster index is selected for the subsequent BP iteration in Steps 10-32. As outlined above, this cluster index depends both on the graph structure and on the received channel values in the previous BP iterations. The output is the reconstructed signal $\hat{\mathbf{x}}$ obtained after executing at most $I_{\max}$ decoding iterations, or until the stopping condition shown in Step 33 is reached. Once decoding ends, we obtain the fully reconstructed signal estimate $\hat{\mathbf{{x}}}=[\hat{x}_0,\ldots,\hat{x}_{n-1}]$.

{{\linespread{0.9}\selectfont  
\begin{algorithm}
%\small
\caption{Learning-based Sequential BP Decoding Scheme}
\SetAlgoLined
\DontPrintSemicolon
\SetKwInOut{Input}{Input}
\SetKwInOut{Output}{Output}
%\underline{function} $\text{IPA}(\boldsymbol{y},\mathbf{A})$\;
\Input{channel information $\mathbf{L}$, parity-check matrix $\mathbf{H}$} %channel information 
\Output{reconstructed signal $\mathbf{\hat{x}}$ }
\label{alg:RL-SD}

Initialization: \;
\hspace{0.25cm} $I\leftarrow 0$\;
\hspace{0.25cm} $m_{c\rightarrow v}\leftarrow 0$ \tcp*{for all CN to VN messages}
\hspace{0.25cm} $m_{v\rightarrow c}\leftarrow L_v$ \tcp*{for all VN to CN messages}
%\hspace{0.25cm} $\hat{\mathbf{L}}_I\leftarrow \mathbf{L}$\;

\If{decoder iteration $I<I_{\max}$} { %stopping condition not satisfied { and} 
\ForEach{cluster with index $a_i$} {
	Determine state $s_{a_i}^{(I)}$\;
}
%{call Algorithm \ref{alg:mrl1_ldpc3} during online local policy adaptation, skip otherwise }\;
{schedule cluster $a_i$ according to policy $\hat{\pi}(s_{a_i}^{(I)})$}\; %for RELDEC, and $\hat{\pi}_k(s_{a_i}^{(I)})$ for AM-RELDEC}\; %Incorporate an optimized cluster scheduling policy\;
\For{selected cluster with index $a_i$} {
		 
	\tcp{decode cluster via flooding}
			\ForEach{CN $c$ in cluster $a_i$} {
		%compute  and propagate $m_{c\rightarrow v}$\;
		%sample  and propagate $m_{v\rightarrow c}$\; %\tcp*{TS step}
		\ForEach{VN $v\in \mathcal{N}(c)$}{
				
			{compute according to (\ref{eq:mcv}) and propagate $m_{c\rightarrow v}^{(I)}$}\;
		}
	}
		
	\ForEach{VN $v$ in the subgraph of cluster $a_i$} {
		\ForEach{CN $c\in \mathcal{N}(v)$}{
				
			{compute according to (\ref{eq:mvc}) and propagate $m_{v\rightarrow c}^{(I)}$}\;
		}

	$\hat{L}_I^{(v)}\leftarrow \sum_{c\in \mathcal{N}(v)} m_{c\rightarrow v}^{(I)}+L_v$ \tcp*{update posterior LLR}
}

\tcp{hard-decision step}
		\ForEach{VN $v$ in the subgraph of cluster $a_i$}{
			\If{$\hat{L}_I^{(v)}\geq 0$}{ 
				$\hat{x}_{v,a_i}^{(I)}\leftarrow 0$\; %tcp*{i.e. use (\ref{eq:pi_G}), (\ref{eq:pi_Q})}
	}
\Else{$\hat{x}_{v,a_i}^{(I)}\leftarrow 1$\;
}
	}

	$i\leftarrow i+1$\;
	}		
	\If{$\mathbf{H}\mathbf{\hat{x}}=\mathbf{0}$}{ 
			break\tcp*{stopping condition reached}
	}
$I\leftarrow I+1$\;
}
\end{algorithm}
}

This learning-based sequential decoding scheme can be viewed as a sequential generalized LDPC (GLDPC) decoder when $z>1$, where BP decoding of a cluster-induced subgraph is analogous to decoding a CN subcode of a GLDPC code. When $z=1$, each cluster represents a single parity-check code, as is the case in a standard  LDPC code. Since the full LDPC Tanner graph is connected and contains cycles, there exist dependencies between the messages propagated by the different clusters of the LDPC code. Consequently, the output of a cluster may depend on messages propagated by previously scheduled clusters. Thus, to improve RL performance for $z>1$, we ensure that the clusters are {chosen to be} as independent as possible. The choice of clustering is determined prior to learning using the cycle-maximization method {discussed in our previous work \cite{HBJNIPS, JSAIT} and omitted here for space considerations.} In short, clusters are selected to maximize the number of cycles in the cluster-induced subgraph to minimize inter-cluster dependencies.

{The principal activity of RELDEC is the computation of the Q-table. Therefore, we are interested to determine how the number of Q-table updates, according to (\ref{eq:q_cls2}), scale with the size of the training dataset and the number of learning steps for a given training sample. Based on this criterion, Algorithm \ref{alg:Qlrn} reveals that the training complexity scales as $\mathcal{O}(|\mathscr{\hat{L}}|\ell_{\max})$.} 

%In the next section, we discuss meta learning variants of RELDEC for obtaining this optimized scheduling policy.

{\section{Channel State Adaptability Via Meta-learning}
\label{sec:meta-RL}
\subsection{Overview}
In order to make RELDEC useful in a practical communication scenario, we need to learn the CN scheduling policy in (\ref{eq:pi_i2}) for a mixture of SNR values. However, it is important to note that such an adaptation can only be done for a limited set of channel SNR values, as the number of elements in this set affects training complexity and the average decoder performance for each specific SNR. This limitation becomes especially relevant in the case of a wireless communication scenario where channel SNR values are frequently changing due to mobility and interference. To address this challenge, we require a decoder that can quickly adapt to the current channel state, \emph{i.e.}, the instantaneous SNR value. In this section, we demonstrate how this can be achieved through meta-RL.}

%where for the case where the current SNR value is nothere is a mismatch between actual and %trained SNRFor example, suppose that each $\mathbf{L}\in \mathscr{\hat{L}}$ in Step 2 of %Algorithm \ref{alg:Qlrn} corresponds to an SNR (in dB) chosen from a set  $\mathcal{S}=%%\{1,1.5,2.5,3,3.5\}$. Now, suppose that during inference, the $\mathbf{L}$ vector input to Algorithm \ref{alg:RL-SD} is based on an SNR value of $1.25$, which is not in $S$. This mismatch of SNR between learning and inference renders (\ref{eq:pi_i2}) sub-optimal for applications in which the channel SNR changes sporadically. Examples of such technologies include IoT based smart transportation, wearable technologies such as fitness trackers, mobile air quality monitoring systems, etc. To improve LDPC decoding performance in these applications, a learning scheme is required which adapts the CN scheduling policy in real-time to the actual channel state, \emph{i.e.,} an SNR of $1.25$. In this section, we show that this task can be accomplished using meta-RL. \\

{A general meta-RL strategy can be outlined as follows. Assume that we have $K$ different tasks to learn. The optimal task-specific policy is learned in two phases: the meta-training phase and the meta-testing (or adaptation) phase. In the meta-training phase, a global long-term expected reward $Q(s,a)$ is learned, as described in Section~\ref{sec:RL}, using a training set derived from a mixture of tasks. The adaptation to the $k$-th task occurs in the meta-testing phase, where $Q(s,a)$ serves as the initialization. From a few additional task-specific training examples, a task-specific policy $\pi_k(s)=\argmax_{a} Q^{(k)}(s,a)$, $k\in[[K]]$ is learned. Here, $Q^{(k)}(s,a)$ represents the task-specific long-term expected reward learned in the adaptation phase. Typically, this adaptation takes place offline (see \cite{Fakoor}).}

{The general structure of our proposed meta-RELDEC scheme is illustrated in Fig.~\ref{fig:meta}. Initially, meta-training is conducted for a set of $K$ different channel SNRs. This process involves learning a global long-term expected reward or action value function, $Q(s_a^{(\ell)},a)$, using LLR vectors corresponding to a mixture of these SNR values. Next, a local long-term expected reward or action value function, $Q^{(k)}(s_a^{(\ell)},a)$, is learned using LLR vectors corresponding to the $k$-th SNR value. These local action values contribute to the enhancement of the global action value $Q(s_a^{(\ell)},a)$ in  subsequent rounds of meta-training. After several iterations, the result of these meta-training rounds  is an optimized global policy $Q^{\ast}(s_a^{(\ell)},a)$. The adaptation to the $k$-th SNR value is then achieved in an online fashion, based on only a few additional training LLR vectors. In a wireless communication context, these LLRs can be easily obtained from received channel pilot symbols, making them readily available. This leads to an SNR-specific scheduling policy. When the channel SNR changes again, the adaptation process restarts from the beginning.}

%During the adaptation phase to the $k$-th SNR value, the global action value function is %used as initialization.
%. Similarly, in case of graph-based decoding (see Fig. \ref{fig:meta}(b)), we learn a %global action value function, $Q^*(s_a^{(\ell)},a)$, during the meta-training phase using %LLR vectors corresponding to a mixture of SNRs, and a local action value function, %$Q^{(k)*}(s_a^{(\ell)},a)$, is learned using LLR vectors corresponding to $k${-th} SNR value %during the adaptation phase, where $k\in[[K]]$. The local and global policies are learned %interactively via offline meta-learning. The resulting global CN scheduling policy is %then used for fast adaptation to a new LLR vector $\mathbf{L}$, corresponding to the %$k$th new SNR, via online meta-learning, which is a novel contribution by itself to the %best of our knowledge. During decoder iteration $I$, the meta-learned $k$th local policy, %$\pi_k^*(s_a^{(I)})$, is invoked to generate an optimal CN scheduling order for %$\mathbf{L}$, resulting in a potentially accurate reconstructed signal $\mathbf{\hat{x}}$.

%The local and global policies are learned interactively via offline meta-learning. 

\vspace{-0.25cm}
\begin{figure}[h]
  \centering
  \includegraphics[scale=0.9]{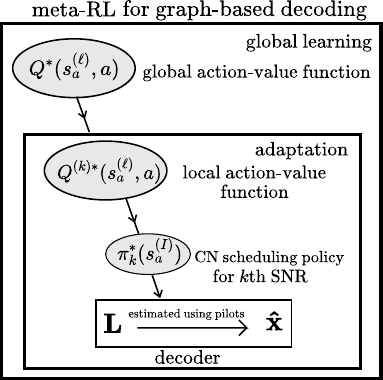}
  %\vspace{-0.5cm}
  \caption{{Meta-learning framework illustration for sequential LDPC decoding. The optimal global action value function $Q^(s_a^{(\ell)},a)$ is learned from a dataset of LLR vectors corresponding to a mixture of SNRs during meta-training. In the adaptation phase, this function is used to learn the local action value function $Q^{(k)}(s_a^{(\ell)},a)$ for the $k$-th local SNR, $k\in [[K]]$, using a small set of additional LLR vectors. The resulting optimal local CN scheduling policy $\pi_k^*(s_a^{(I)})$ is then applied for sequential LDPC decoding. }}
  \label{fig:meta}
\end{figure}
{\subsection {Meta Reinforcement Learning for Optimal CN Scheduling}
In the following, we present a meta-RL method that directly estimates the Q-function by minimizing a Q-learning-based loss function over mini-batches of MDP instances. This approach enables fast and efficient adaptation to varying channel SNRs by terminating the meta-learning algorithm when the loss is below a certain threshold. We further discuss the requirements for successful Q-learning, the need for a large set of channel information vectors, and the optimization of global and local scheduling policies. }
{Recall that the RL environment is modeled as a finite MDP, which can be seen as a sequence of state, action, and reward transitions (see Fig. \ref{fig:RL}). These transitions can be expressed as a tuple $(s_a^{(\ell)},a,R_a,s_a^{(\ell)'})$ specific to each learning step $\ell$. Throughout the paper, we refer to this tuple as an \emph{MDP instance}. Let $\mathcal{D}({\pi(s_a^{(\ell)}),\mathbf{L}})\triangleq (s_{a_0}^{(0)},{a_0},R_{a_0},s_{a_0}^{(0)'}),\ldots,(s_{a_{\ell_{\max}-1}}^{(\ell_{\max}-1)},a_{\ell_{\max}-1}, R_{a_{\ell_{\max}-1}},$ $s_{a_{\ell_{\max}-1}}^{(\ell_{\max}-1)'})$ represent a \emph{mini-batch} of $\ell_{\max}$ MDP instances. These instances are obtained after taking $\ell_{\max}$ actions $a_0,\ldots,a_{\ell_{\max}-1}$ according to policy $\pi(s_a^{(\ell)})$, $a\in [[\ceil{m/z}]]$, where $s_{a_0}^{(0)}$ is the initial state of the MDP after observing $\mathbf{L}$ during a learning episode. For instance, if $\ell_{\max}=2$, $a_0=1$, and $a_1=2$, the second and third clusters are scheduled in the first two learning steps, and $\mathcal{D}({\pi(s_a^{(\ell)}),\mathbf{L}})= {(s_{1}^{(0)},{1},R_{1},s_{1}^{(0)'}),(s_{2}^{(1)},{2},R_{2},s_{2}^{(1)'})}$.}
{In comparison to MAML, which uses gradient descent to optimize the model parameters  \cite{Finn}, our meta-RL scheme directly estimates the Q-function by minimizing a Q-learning-based loss function over a mini-batch $\mathcal{D}({\pi(s_a^{(\ell)}),\mathbf{L}})$. We will now discuss a method of deriving this function. To do so, note that the Q-learning update procedure shown in (\ref{eq:q_cls2}) can also be expressed as
\begin{align}
	\begin{split}
	&Q_{\ell+1}(s_a^{(\ell)},a)= Q_\ell(s_a^{(\ell)},a)+ \alpha(U_\ell(s_a^{(\ell)},a)-Q_\ell(s_a^{(\ell)},a)),
	\end{split}
	\label{eq:q_clsT}
\end{align}
 where $U_\ell(s_a^{(\ell)},a)=R_a+\beta \max_{a'}Q_\ell(s_a^{(\ell)'},a')$, and $U_\ell(s_a^{(\ell)},a)-Q_\ell(s_a^{(\ell)},a)$ is the temporal difference (TD) error \cite{Sutton18}. As learning iteration $\ell$ increases, $Q_{\ell+1}(s_a^{(\ell)},a)$ approaches $Q^*(s_a^{(\ell)},a)$, \emph{i.e.,} the meta-RL scheme converges to the true Q-function $Q^*(s_a^{(\ell)},a)$ \cite{sutton88}, \cite[Sec. 6.4]{Sutton18}. Consequently, the TD error approaches $0$ as $\ell \rightarrow \infty$. Assume that meta-RL is carried out on mini-batch $\mathcal{D}({\pi(s_a^{(\ell)}),\mathbf{L}})$ during a learning episode. At any given learning step $\ell$, the meta-RL algorithm computes a sum of squared TD errors over this mini-batch, given by
\begin{equation}
\begin{split}
\label{eq:loss}
\mathcal{L}(\mathcal{D}({\pi(s_a^{(\ell)}),\mathbf{L}}))=\sum_{(s_a^{(\ell)},a,R_a,s_a^{(\ell)'})\in \mathcal{D}({\pi(s_a^{(\ell)}),\mathbf{L}})}&(U_{\ell}(s_a^{(\ell)},a)- \\
&Q_{\ell}(s_a^{(\ell)},a))^2,
\end{split}
\end{equation}
which is minimized as the agent takes an action via an $\epsilon$-greedy policy at each learning step. Calculating $\mathcal{L}(\mathcal{D}({\pi(s_a^{(\ell)}),\mathbf{L}}))$ enables us to terminate the meta-learning algorithm once the loss falls below a certain threshold, resulting in a fast and efficient adaptation phase. This distinguishing feature sets our proposed meta-learning approach apart from the baseline RELDEC scheme presented in Section \ref{sec:RL}. }
{It is important to note that a mini-batch $\mathcal{D}({\pi(s_a^{(\ell)}),\mathbf{L}})$ contains MDP instances resulting from a single channel output $\mathbf{L}$. However, successful Q-learning requires a substantial amount of training data, and the MDP instances derived from a single training sample does not provide sufficient exploration for the agent. To address this issue,  we perform meta-RL on an adequately large set of channel information (LLR) vectors $\mathscr{L}=\{\mathbf{L}_0,\ldots,\mathbf{L}_{|\mathscr{L}|-1}\}$. Hence, in each episode of the proposed meta-RL algorithm, the Q-learning loss is calculated  over a batch $\mathcal{B}({\pi(s_a^{(\ell)}),\mathscr{L}})\triangleq \{\mathcal{D}({\pi(s_a^{(\ell)}),\mathbf{L}_0}),\ldots,$ $\mathcal{D}({\pi(s_a^{(\ell)}),\mathbf{L}_{|\mathscr{L}|-1}})\}$ of MDP instances, which consists of  $|\mathscr{L}|$ mini-batches. }
{The corresponding batch loss $\mathcal{L}(\mathcal{B}({\pi(s_a^{(\ell)}),\mathscr{L}}))$ is an aggregate of all mini-batch losses \\ $\mathcal{L}(\mathcal{D}({\pi(s_a^{(\ell)}),\mathbf{L}_0})),\ldots,\mathcal{L}(\mathcal{D}({\pi(s_a^{(\ell)}),\mathbf{L}_{|\mathscr{L}|-1}}))$,  generated by the $|\mathscr{L}|$ learning episodes. The loss must be minimized to ensure Q-learning convergence. In the following, we will discuss how the output of this loss function is minimized in each local and global learning phase of meta-learning.}
	
%, in addition to $\mathcal{L}(\mathcal{D}({\pi(s_a^{(\ell)}),\mathbf{L}}))$

{Let $\mathcal{B}({\pi(s_a^{(\ell)}),\mathscr{L}})$ (resp. $\mathcal{B}_k({\pi_k(s_a^{(\ell_k)}),\mathscr{L}_k})$) denote a batch of MDP instances used for learning the global (resp., local) CN scheduling policy. Here, $\mathscr{L}$ contains LLR vectors corresponding to a mixture of $K$ distinct SNR values, \emph{i.e.,} there are $K$ subsets in $\mathscr{L}$, each containing $|\mathscr{L}|/K$ LLR vectors corresponding to a specific SNR. On the other hand, all LLR vectors in $\mathscr{L}_k$ correspond to a single SNR value. The global optimization problem solved by Q-learning can be described by finding the best global action value (or long-term expected reward) function which minimizes the expected loss. By using \eqref{eq:loss}, this can be expressed as}
%, learned by observing all the MDP instances in $\mathcal{B}({\pi(s_a^{(\ell)}),\mathscr{L}})$, %is expressed by using \eqref{eq:loss} as}
%
%\vspace{-0.5cm}
\begin{align}
\label{eq:opt1}
\begin{split}
	Q^*(s_a^{(\ell)},a)&=\argmin_{Q_\ell(s_a^{(\ell)},a)\in \mathbb{R}} \mathbb{E}_{\mathbf{L}\in \mathscr{L}}[\mathcal{L}(\mathcal{D}({\pi(s_a^{(\ell)}),\mathbf{L}}))]\\
	&=\argmin_{Q_\ell(s_a^{(\ell)},a)\in \mathbb{R}} \mathbb{E}_{\mathbf{L}\in \mathscr{L}}\bigg[ \sum_{(s_a^{(\ell)},a,R_a,s_a^{(\ell)'})\in \mathcal{D}({\pi(s_a^{(\ell)}),\mathbf{L}})} \\
	&(U_{\ell}(s_a^{(\ell)},a)-Q_{\ell}(s_a^{(\ell)},a))^2 \bigg].
	\end{split}
\end{align}
%where $\mathbf{l}$ is a realization of $\mathbf{L}$. 
{The optimal global CN scheduling policy, $\pi^*(s_{a_i}^{(I)})$, obtained at the end of the global learning phase is defined as	}
\begin{equation}
	\label{eq:exp2b}
	\pi^*(s_{a_i}^{(I)})\triangleq \argmax_{a_i\in [[\ceil{m/z}]]\setminus \{a_0,\ldots, a_{i-1}\}} Q^*(s_{a_i}^{(I)},a_i),
\end{equation} 
{where $I$ is the decoder iteration during inference. }

{Likewise, finding the best  $k$-th \emph{local} action value or reward function for the $k$-th SNR value can be described by the following optimization problem:} %learned by observing all the MDP instances in $\mathcal{B}_k({\pi_k(s_a^{(\ell)}),\mathscr{L}_k})$ is expressed as the argmin over all action value functions $Q^{(k)}(s_a^{(\ell)},a)$ for the $k$-th SNR
%\vspace{-0.5cm}
%{, and $Q^*(s_{a_i}^{(I)},a_i)$ is learned using the dataset (set of LLR vectors) corresponding to a mixture of SNRs
\begin{align}
\begin{split}
\label{eq:opt2}
	Q^{(k)*}(s_a^{(\ell)},a)=\argmin_{Q_\ell^{(k)}(s_a^{(\ell)},a)\in \mathbb{R}} \mathbb{E}_{\mathbf{L}'\in \mathscr{L}_k}[\mathcal{L}(\mathcal{D}_k({\pi_k(s_a^{(\ell)}),\mathbf{l}'}))],\\
	\quad k\in[[K]].
\end{split}
\end{align}
{ Here, $\mathbf{L}'\in \mathscr{L}_k=\{\mathbf{L}_0^{(k)'},\ldots,$ $\mathbf{L}_{|\mathscr{L}_k|-1}^{(k)'}\}$ is an LLR vector taken from the set  $\mathscr{L}_k$ for the $k$-th SNR value and $\mathcal{D}_k({\pi_k(s_a^{(\ell)}),\mathbf{L}'})$ is the corresponding  mini-batch of MDP instances for  $k\in[[K]]$.}
% Similiarly to the global policy, the $k$-th local policy is determined as
%\vspace{-0.5cm}
%\begin{equation}
%	\label{eq:pi_Qk2}
%	\pi_k(s_a^{(\ell)})=\argmax_{a} Q^{(k)}(s_a^{(\ell)},a),
%\end{equation}
%where $Q^{(k)}(s_a^{(\ell)},a)$ represents some action value function corresponding to the $k$-th %policy. 
{Similarly to the global policy, the  optimal $k$-th local CN scheduling policy $\pi^*_k(s_{a_i}^{(I)})$ used for inference is then given as} 
%\vspace{-0.5cm}
\begin{equation}
	\label{eq:exp3}
	\pi^*_k(s_{a_i}^{(I)})\triangleq \argmax_{a_i\in [[\ceil{m/z}]]\setminus \{a_0,\ldots, a_{i-1}\}} Q^{(k)*}(s_{a_i}^{(I)},a_i),
\end{equation}
{where $Q^{(k)*}(s_a^{(I)},a)$ is given by \eqref{eq:opt2} for scheduling the $i$-th CN $a_i$ in each cluster.}

{In the remainder of the paper, we use simplified notations $\mathcal{B}$ and $\mathcal{B}_k$ to represent the MDP batches $\mathcal{B}({\pi(s_a^{(\ell)}),\mathscr{L}})$ and $\mathcal{B}_k({\pi_k(s_a^{(\ell)}),\mathscr{L}_k})$, respectively. Additionally, we use simplified notations $\mathcal{D}_{\mathbf{L}}$ and $\mathcal{D}_{\mathbf{L}'}$ for $\mathcal{D}(\pi(s_a^{(\ell)}),$ $\mathbf{L})$ and $\mathcal{D}_k({\pi_k(s_a^{(\ell)}),\mathbf{L}'})$. Since searching over all possible action values for a given $(s_a^{(\ell)},a)$ pair in (\ref{eq:opt1}) and (\ref{eq:opt2}) is computationally prohibitive, an alternative approach is to employ standard Q-learning. This method yields optimized global and $k$-th local CN scheduling policies by iteratively minimizing empirical losses $\mathcal{L}=\frac{1}{|\mathcal{B}|} \mathcal{L}(\mathcal{B})$ and $\mathcal{L}_k=\frac{1}{|\mathcal{B}_k|} \mathcal{L}(\mathcal{B}_k)$, respectively, during each learning episode. }

{Subsequently, a global (resp., local) policy update refers to learning the CN scheduling policy by minimizing  $\mathcal{L}$ (resp., $\mathcal{L}_k$). Although the distribution of the reward $R_a$ for scheduling cluster $a$ may differ across tasks, we consider the tasks related since the environment (the sequential BP decoder) remains unchanged as learning progresses. In the following, we introduce our meta-RL scheme AM-RELDEC, which involves a bi-level policy optimization, as a novel extension of the MAML scheme \cite{Finn} applied to Q-learning. This approach learns a global CN scheduling policy using a dataset corresponding to an SNR mixture and $K$ local CN scheduling policies based on $K$ separate datasets corresponding to individual SNRs, interactively.} 

{\subsection{AM-RELDEC Algorithm: Learning the Global Policy}
This subsection introduces the AM-RELDEC scheme as shown in Algorithm \ref{alg:mrl1_ldpc}. This novel meta-learning scheme, which has not been published in the open literature to our knowledge, is well suited for wireless communications scenarios with varying channel conditions due to its agility. The global policy for CN scheduling can rapidly adapt online to any local policy corresponding to a particular SNR during the decoding phase. The scheme takes $\mathscr{L}$, $\mathscr{L}_k$, and $\mathbf{H}$ as inputs and outputs an optimized global policy, which serves as a starting point for optimizing the $k$-th local policy during online adaptation, $k\in[[K]]$.}

%The AM-RELDEC scheme is shown in Algorithm \ref{alg:mrl1_ldpc}. It takes $\mathscr{L}$, %$\mathscr{L}_k$ and $\mathbf{H}$ as inputs, and outputs an optimized global policy, $\hat{\pi}(s_{a_i}^{(I)})$, which is used as a starting point for optimizing the $k$-th local policy during online adaptation. 
{Initially, the $K$ local learning stages of AM-RELDEC are completed in Steps 4-31. In Step 5, the agent initializes a local CN scheduling policy $\pi_k(s_a^{(\ell)})$ using a global CN scheduling policy $\pi(s_a^{(\ell)})$ at the beginning of the $k$-th adaptation phase. Between Steps 6-28, the agent then optimizes this policy by minimizing the sum of squared TD errors for batch $\mathcal{B}_k$ of MDPs, taking actions according to an $\epsilon$-greedy policy in Step 10}:
 \begin{equation}
	\label{eq:egreedy2}
	a=
	\begin{cases}
     \text{selected uniformly at random w.p. } \epsilon \text{ from }[[\ceil{m/z}]], \\
	  f(s_a^{(\ell)}) \text{ selected w.p. } 1-\epsilon, 
	\end{cases}
\end{equation} 
{where $f(s_a^{(\ell)}) =\pi_k(s_a^{(\ell)})$. For global policy optimization, the agent {invokes Algorithm \ref{alg:mrl1_ldpc2} in Step 32. In Step 1 of this algorithm, the agent initializes $\pi(s_a^{(\ell)})$ using the learned local policies and  optimizes it by minimizing the squared TD error for the global batch $\mathcal{B}$ in Steps 2-13. This procedure is repeated in every meta-learning iteration {starting in Step 2 of Algorithm \ref{alg:mrl1_ldpc}}. Upon completion of learning, an optimized version of the global policy $\hat{\pi}(s_{a_i}^{(I)})$  and an optimized version of the $k$-th local policy  $\hat{\pi}_k(s_{a_i}^{(I)})$, are obtained.}}

{\linespread{1}\selectfont  
\begin{algorithm}
%\small
\caption{AM-RELDEC} %AM-RELDEC for Sequential LDPC Decoding
\SetAlgoLined
\DontPrintSemicolon
\SetKwInOut{Input}{Input}
\SetKwInOut{Output}{Output}
%\underline{function} $\text{IPA}(\boldsymbol{y},\mathbf{A})$\;
\Input{set of LLR vectors $\mathscr{L}$, $\mathscr{L}_k$, parity-check matrix $\mathbf{H}$}
%\Output{meta-learning loss $\mathcal{L}(\mathcal{D})$ } 
\Output{optimized global scheduling policy $\hat{\pi}(s_{a_i}^{(I)})$}
\label{alg:mrl1_ldpc}

$U_0(s_a^{(0)},a) \leftarrow 0$, $Q_0(s_a^{(0)},a) \leftarrow 0$, $\forall$ $s_a^{(0)},a$, $\mathcal{B}\leftarrow \emptyset$
\hspace{0.25cm}

\While{not done} {
\tcp{meta-learning phase}
$k\leftarrow 1$\; 
\While{$k\leq K$} {
	\tcp{adapt to the $k$-th SNR}
	$\mathcal{B}_k\leftarrow \emptyset$, $Q^{(k)}(s_a^{(0)},a) \leftarrow Q_0(s_a^{(0)},a)$ $\forall$ $s_a^{(0)},a$\;
	\tcp{start of an episode}
	\For{each new $\mathbf{L}'\in \mathscr{L}_k$} {
		$\ell\leftarrow 0$, $\mathbf{\hat{L}}_{\ell}\leftarrow \mathbf{L}'$, $\mathcal{D}_{\mathbf{L}'}\leftarrow \emptyset$, $\mathcal{L}_k\leftarrow 1$\;
		determine initial states of all clusters using (\ref{eq:hd})\;
		\While{$\mathcal{L}_k> \mathcal{L}_{\min}$ and $\ell<\ell_{\max}$} {
			select cluster $a$ according to {(\ref{eq:egreedy2})}\; %$\epsilon$-greedy policy $\pi_Q^{(\ell)}$		 
			% {Step 8} of Algorithm \ref{alg:Qlrn}\;
			{decode cluster induced subgraph according to Steps 8-9 of Algorithm \ref{alg:Qlrn}\;}
			{determine $\mathbf{\hat{x}}_a$ using Steps 10-17 of Algorithm \ref{alg:Qlrn}\;}
		determine index $s_a^{(\ell)'}$ of $\mathbf{\hat{x}}_a$ via binary to decimal conversion\;
		update $R_a$ according to (\ref{eq:rew})\;
		$U_\ell(s_a^{(\ell)},a) \leftarrow R_a+\beta \max_{a'\in [[\ceil{m/z}]]}Q_\ell(s_a^{(\ell)'},a')$\;
		compute $Q_{\ell+1}(s_a^{(\ell)},a)$ according to (\ref{eq:q_cls2})\;
		$s_a^{(\ell+1)}\leftarrow s_a^{(\ell)'}$\;
		$\mathcal{D}_{\mathbf{L}'}\leftarrow \mathcal{D}_{\mathbf{L}'} \cup (s_a^{(\ell)},a,R_a,s_a^{(\ell)'})$\;
		\For{every $x$ new MDP instances in $\mathcal{D}_{\mathbf{L}'}$}{
		$\mathcal{L}(\mathcal{D}_{\mathbf{L}'})\leftarrow \sum_{(s_a^{(\ell)},a,R_a,s_a^{(\ell)'})\in \mathcal{D}_{\mathbf{L}'}}(U_{\ell}(s_a^{(\ell)},a)-Q_{\ell}(s_a^{(\ell)},a))^2$\;
		$\mathcal{B}_k\leftarrow \mathcal{B}_k \cup \mathcal{D}_{\mathbf{L}'}$\;
	$\mathcal{L}(\mathcal{B}_k)\leftarrow \mathcal{L}(\mathcal{B}_k) + \mathcal{L}(\mathcal{D}_{\mathbf{L}'})$\;
		$\mathcal{L}_k\leftarrow \frac{1}{|\mathcal{B}_k|} \mathcal{L}(\mathcal{B}_k) $ \tcp{local error minimized as learning continues}
		}
			$\ell\leftarrow \ell+1$\; 
		}
		$Q_0(s_a^{(\ell)},a) \leftarrow Q_\ell(s_a^{(\ell)},a)$ $\forall$ $s_a^{(\ell)},a$\;
}
	$Q^{(k)}(s_a^{(\ell)},a) \leftarrow Q_0(s_a^{(\ell)},a)$ $\forall$ $s_a^{(\ell)},a$ \tcp{updates the $k$-th local policy} % of (\ref{eq:pi_Qk2})
	$k\leftarrow k+1$\; 
}
perform Steps 1-13 of Algorithm \ref{alg:mrl1_ldpc2} \tcp{global policy update}
}	
\end{algorithm}
}

{\linespread{0.99}\selectfont  
\begin{algorithm}[h]
%\small
\caption{AM-RELDEC (continued from Step 32 of Algorithm \ref{alg:mrl1_ldpc})}
\SetAlgoLined
\DontPrintSemicolon
\label{alg:mrl1_ldpc2}

$Q_0(s_a^{(0)},a) \leftarrow \frac{1}{K}\sum_{k=1}^{K}Q^{(k)}(s_a^{(0)},a)$ $\forall$ $s_a^{(0)},a$ \tcp{initializes $\pi(s_a^{(\ell)})$} % of (\ref{eq:pol_clus})
%perform Steps 3-19 of Algorithm \ref{alg:mrl2}\;
\tcp{start of an episode}
\For{each new $\mathbf{L}\in \mathscr{L}$} {
	$\ell\leftarrow 0$, $\mathbf{\hat{L}}_{\ell}\leftarrow \mathbf{L}$, $\mathcal{D}_{\mathbf{L}}\leftarrow \emptyset$\;
\While{$\ell<\ell_{\max}$}
	{
		select cluster $a$ according to \ref{eq:egreedy2})\;
		repeat Steps 11-17 of Algorithm \ref{alg:mrl1_ldpc}\;
		$\mathcal{D}_{\mathbf{L}}\leftarrow \mathcal{D}_{\mathbf{L}} \cup (s_a^{(\ell)},a,R_a,s_a^{(\ell)'})$\;
		$\ell\leftarrow \ell+1$\; %tcp
	}
	$\mathcal{L}(\mathcal{D}_{\mathbf{L}})\leftarrow \sum_{(s_a^{(\ell_\max)},a,R_a,s_a^{(\ell_\max)'})\in \mathcal{D}_{\mathbf{L}}}(U_{\ell_\max}(s_a^{(\ell_\max)},a)-Q_{\ell_\max}(s_a^{(\ell_\max)},a))^2$\;
	$\mathcal{B}\leftarrow \mathcal{B} \cup \mathcal{D}_{\mathbf{L}}$\;
	$Q_0(s_a^{(\ell_\max)},a) \leftarrow Q_{\ell_\max}(s_a^{(\ell_\max)},a)$ $\forall$ $s_a^{(\ell_\max)},a$\; %, U_0(s,a) \leftarrow U_{\ell_\max}(s,a)
$\mathcal{L}\leftarrow \frac{1}{|\mathcal{B}|} \mathcal{L}(\mathcal{B})$ \tcp{global error minimized as learning continues}
}
	
\end{algorithm}
}

{Learning is carried out through a full BP iteration, as demonstrated in Step 11, based on the chosen cluster $a$ in Step 10 of Algorithm \ref{alg:mrl1_ldpc}. The Q-learning error $\mathcal{L}(\mathcal{D}_{\mathbf{L}'})$ is updated in Step~20 after every $x\ll \ell_{\max}$ training steps in each learning episode, where a mini-batch $\mathcal{D}_{\mathbf{L}'}\subset \mathcal{B}_k$ contains MDP instances corresponding to a training sample $\mathbf{L}'$. This update  may result in an overall loss $\mathcal{L}_k$ that is smaller than a threshold $\mathcal{L}_{\min}$. If this occurs, learning can proceed to the next training example before completing all $\ell_{\max}$ learning steps for the current example. The local policy $\pi_k(s_a^{(\ell)})$, optimized in Step 29 as $Q_\ell(s_a^{(\ell)},a)$, is iteratively updated in Step 16 by taking actions in Step 10 according to (\ref{eq:egreedy2}). These Q-learning updates decrease the local loss $\mathcal{L}_k$ in Step 23.}

{The global policy for a mixture of $K$ SNRs is optimized in Steps 2-{13} of Algorithm \ref{alg:mrl1_ldpc2} (called from Step {32} of Algorithm \ref{alg:mrl1_ldpc}). In Step 1, the action value function corresponding to the global policy $\pi(s_a^{(\ell)})$ is determined by calculating the average action values across all $K$ local action value functions. In each learning  episode, global learning undergoes $\ell_{\max}$ training rounds as shown in Steps 4-9. Consequently, for the training example $\mathbf{L}\in \mathscr{L}$, the cardinality of the corresponding mini-batch $\mathcal{D}_{\mathbf{L}} \subset \mathcal{B}$ is $\ell_{\max}$. As the agent continously interacts with the environment, the global action value $Q_{\ell+1}(s_a^{(\ell)},a)$ is updated,  and $\pi(s_a^{(\ell)})$ is optimized by taking actions according to (\ref{eq:egreedy2}), where $f(s_a^{(\ell)}) =\pi(s_a^{(\ell)})$, This process leads to gradual reduction of the global loss $\mathcal{L}(\mathcal{B})$. Once all meta-iterations are completed, Algorithm \ref{alg:mrl1_ldpc} generates an optimized global policy $\hat{\pi}(s_{a_i}^{(I)})$.  Note that the complexity of AM-RELDEC, as seen from Algorithms \ref{alg:mrl1_ldpc} and \ref{alg:mrl1_ldpc2} is $\mathcal{O}((|\mathscr{L}|+K|\mathscr{L}_k|)\ell_{\max})$. }

% by minimizing $\mathcal{L}(\mathcal{B})$

{\linespread{0.99}\selectfont  
\begin{algorithm}[h]
%\small
\caption{AM-RELDEC (online adaptation phase)}
\SetAlgoLined
\DontPrintSemicolon
\SetKwInOut{Input}{Input}
\SetKwInOut{Output}{Output}
%\underline{function} $\text{IPA}(\boldsymbol{y},\mathbf{A})$\;
\Input{set of LLR vectors $\mathscr{L}_k'$ obtained after channel estimation, parity-check matrix $\mathbf{H}$, action values $Q_0(s_a^{(0)},a)$ corresponding to optimized global policy } %obtained according to (\ref{eq:exp2b})
%\Output{meta-learning loss $\mathcal{L}(\mathcal{D})$ } 
\Output{optimized local scheduling policy $\hat{\pi}_k(s_{a_i}^{(I)})$} % of (\ref{eq:pi_i3})
\label{alg:mrl1_ldpc3}
$\mathcal{B}_k\leftarrow \emptyset$, $Q^{(k)}(s_a^{(0)},a) \leftarrow Q_0(s_a^{(0)},a)$ $\forall$ $s_a^{(0)},a$\;
\tcp{start of an episode}
\For{each new $\mathbf{L}'\in \mathscr{L}_k'$} {
	perform Steps 7-27 of Algorithm \ref{alg:mrl1_ldpc} \tcp{adapt to the current SNR with index $k$}
}
$Q^{(k)}(s_a^{(\ell)},a) \leftarrow Q_0(s_a^{(\ell)},a)$ $\forall$ $s_a^{(\ell)},a$\; % \tcp{updates the $k$-th policy of (\ref{eq:pi_Qk2})}	
{$\pi_k(s_a^{(\ell)})\leftarrow \argmax_{a} Q^{(k)}(s_a^{(\ell)},a)$}\tcp{$k$-th local policy update}
\end{algorithm}
}

{\subsection{AM-RELDEC Algorithm: Inference and Online Adaptation}
Assume that a global policy for a mixture of $K$ SNR values has been already learned through Algorithms \ref{alg:mrl1_ldpc} and \ref{alg:mrl1_ldpc2} and stored at the decoder. Now, at decoding time, assume that the $k$-th SNR, $k\in[[K]]$, is observed at the channel output.  This section addresses how, at decoding time, an optimal CN scheduling is obtained by online adaptation to the observed SNR value. The global policy is used to initialize the online adaptation.  We focus on  a wireless communication setting where the channel is estimated accurately using pilot signals. The  estimation results in a set of LLR vectors $\mathscr{L}_k'$, with each vector $\mathbf{L}'\in \mathscr{L}_k'$ corresponding to a new SNR value. The SNR values are used to retrain the scheduling policy in the online adaptation shown in Algorithm \ref{alg:mrl1_ldpc3}.}

{This algorithm takes as inputs the LLR values in $\mathscr{L}_k'$. The other input  comprises the action values $Q_0(s_a^{(0)},a)$ for all $s_a^{(0)},a$, which are set according to the global policy learned earlier. In Step 1 of Algorithm \ref{alg:mrl1_ldpc3}, local action values $Q^{(k)}(s_a^{(0)},a)$ for all $s_a^{(0)},a$ are initialized using the global action values. Steps 2-4 update the local policy with all the training samples in $\mathscr{L}_k'$. In particular, Step 3 adapts the $k$-th local policy using Steps 7-27 of Algorithm \ref{alg:mrl1_ldpc}. Note, when $\mathcal{L}_k \leq \mathcal{L}_\min$ in Step 9 of Algorithm \ref{alg:mrl1_ldpc}, an episode terminates before completing all $\ell_\max$ iterations, accelerating the online adaptation phase.}

The  output of Algorithm \ref{alg:mrl1_ldpc3} is an optimized CN scheduling policy, $\hat{\pi}_k(s_{a_i}^{(I)})$, for a new SNR, given by
% which can be viewed as the final adaptation step of Algorithm \ref{alg:mrl1_ldpc}, \emph{i.e.,} the last meta-iteration is online, and all prior iterations are offline
\begin{equation}
	\label{eq:pi_i3}
 \hat{\pi}_k(s_{a_i}^{(I)})=\argmax_{a_i\in [[\ceil{m/z}]]\setminus \{a_0,\ldots, a_{i-1}\}} Q^{(k)}(s_{a_i}^{(I)},a_i). 
\end{equation} 
This policy is then used to select the CN updates during decoding in the same way as in the RELDEC scheme.
%{This policy is employed} in Step {10} of Algorithm \ref{alg:RL-SD} during %{inference.}

{The iterative learning of the global policy along with the $K$ local policies using Algorithms \ref{alg:mrl1_ldpc} and \ref{alg:mrl1_ldpc2} enables fast adaptation to any new current channel condition through Algorithm~\ref{alg:mrl1_ldpc3} using only a relatively small number of LLR vectors in $\mathscr{L}_k'$.  This is in sharp contrast to the meta Q-learning scheme of \cite{Fakoor}, where a local policy is learned after  learning the global policy, but  the learned local policies are not employed for global policy updates. As a result, the scheme's global policy is static and unsuitable for online local policy adaptation.}

{It is important to note that the  complexity of Algorithm \ref{alg:mrl1_ldpc3} scales as $\mathcal{O}(|\mathscr{L}_k'|\ell_{\max})$, which is  small compared to  RELDEC's complexity  during inference,  as typically $|\mathscr{L}_k'|\ll |\mathscr{\hat{L}}|$. Additionally, AM-RELDEC eliminates the need to store separate Q-tables for the local policies, which can be considerably large for the codes considered in this work. Instead, only the global action table needs to be stored, which is then adapted online to the actual SNR value of the channel. }

%\vspace{-3ex}
\section{Experimental Results}
%\vspace{-1ex}
\label{sec:results}

In this section, we compare the performance of our learning-based sequential decoding schemes with flooding (\textit{i.e.}, all clusters are updated simultaneously per iteration), and a random sequential decoding scheme where the cluster scheduling order is randomly generated. This is similar to the layered decoding scheme in which the CN scheduling order is fixed \cite{CGW10}. We employ each scheme for decoding a $[384,256]$-WRAN irregular LDPC code (see \cite{wran}), a $(3,5)$ AB-LDPC code of block length $500$ bits, and a 5G-NR LDPC code. For the 5G-NR case, the code is constructed by lifting the BG2 base matrix{\footnote{{The BG2 base matrix has $38$ columns of weight one and is available at \cite{5gbase}.}}} with dimensions $42\times 52$ using an optimized lifting matrix obtained from the literature (see \cite{RK18,Bae}) with a lifting factor $10$, resulting in a 5G-NR LDPC code with block length $520$ and a rate of approximately $1/5$. The simulation of 5G-NR LDPC codes based on the BG1 matrix is beyond the scope of this work, as the graph contains several degree-$19$ CNs, which, once the code is lifted, renders both reinforcement and meta-RL extremely computationally intensive. For all codes, the choice of block length is influenced by the run-time of the proposed learning schemes on our system. We employ both  RELDEC in Algorithm \ref{alg:Qlrn} and its meta-variant AM-RELDEC in Algorithms \ref{alg:mrl1_ldpc} and \ref{alg:mrl1_ldpc3}, respectively.
%38 cols of BG2 has weight one

%{For comparison purposes we also present a RL scheme based on deep learning. Specifically,  for MDPs with very large state spaces, the action value function $Q_\ell(s,a)$ can be approximated as $Q_\ell(s,a;\mathbf{W})$ using a deep learning model with  tensor $\mathbf{W}$ representing the weights connecting all layers in the NN (see e.g. \cite[Chap. 9]{Sutton18}). We utilize a separate NN with weight $\mathbf{W}_\ell^{(a)}$ for each cluster in each learning step $\ell$, since a single NN cannot distinguish between the signals $\mathbf{\hat{x}}_0^{(\ell)},\ldots,\mathbf{\hat{x}}_{\ceil{m/z}-1}^{(\ell)}$, and hence the rewards $R_0,\ldots,R_{\ceil{m/z}-1}$ generated by the $\ceil{m/z}$ different clusters. {Each NN can be viewed as a substitute of a sub-Q-table with rows corresponding to the state indices of a particular cluster.} This scheme is {called} Deep-RELDEC (D-RELDEC) in the following.} 

%A cluster NN learns to map the cluster output $\mathbf{\hat{x}}_a^{(\ell)}$ to a vector of $\ceil{m/z}$ predicted action values $[Q_\ell(s_a',0;\mathbf{W}_\ell^{(a)}), $ $\ldots, Q_\ell(s_a',\ceil{m/z}-1;\mathbf{W}_\ell^{(a)})]$. 

%{Here, each cluster NN is based on a feed-forward architecture with an input layer of size $l_a$, two hidden layers of sizes $250$ and $125$, respectively, and an output layer of size $\ceil{m/z}$. The activation function used for the hidden and output layers are rectified linear unit and sigmoid, respectively.}

After  learning is complete, the resulting cluster scheduling policy for each code is incorporated into Step 9 of Algorithm \ref{alg:RL-SD}, yielding either a a RELDEC or  AM-RELDEC sequential decoding scheme for that code, depending on the chosen learning algorithm. In the case of RELDEC, the LLR vectors used for training are sampled uniformly at random over a range of $K$ equally-spaced SNR values for a given code. As a result, there are $\mathscr{\hat{L}}|/K$ LLR vectors in $\mathscr{\hat{L}}$ for each SNR value considered, meaning that  {RELDEC is trained on a mixture of SNR values}. On the other hand, for AM-RELDEC, there are $|\mathscr{L}|/K$ (resp., $|\mathscr{L}_k|$) LLR vectors in $\mathscr{L}$ (resp., $\mathscr{L}_k$) trained individually for each fixed SNR. 

For all learning schemes, we employ a reasonable choice of hyper-parameters that lead to good decoding performance, such as a learning rate of $\alpha=0.1$, a reward discount rate of $\beta=0.9$, a probability of exploration {of} $\epsilon=0.6$, a maximum number of steps per learning episode of $\ell_{\max}=50$, and a Q-learning loss threshold of $\mathcal{L}_{\min}=1\times 10^{-4}$. For RELDEC and AM-RELDEC utilize a cluster size of $z=1$. Note that the total number of SNR values of $K=6$ is chosen for all codes. Specifically, we consider SNR values of $1, 2,\ldots,5$ and $5.5$ dB for the WRAN, $1, 1.5, \ldots,3$ {and $3.5$} dB for the 5G-NR, and $1, 1.5,\ldots,3$ {and $3.25$} dB for the AB LDPC code.

For RELDEC, we set $|\mathscr{\hat{L}}|=15000$, ensuring that $1/K$-th of the dataset contains LLR vectors of a fixed SNR; \emph{i.e.,} these schemes are  trained for a mixture of $K$ SNR values. On the other hand, for AM-RELDEC, we obtain optimized local policies by performing $100$ meta-iterations, with $99$ of them used for global-policy learning using Algorithm \ref{alg:mrl1_ldpc}, and the last one used for adaptation using Algorithm \ref{alg:mrl1_ldpc3}. We choose $|\mathscr{L}|=7425$, $|\mathscr{L}_k|=7425$, and consequently $75$ training examples are used for the learning the global and $k$-th local policy, respectively, in each meta-iteration. {Additionally, for the online adaptation training set we consider $|\mathscr{L}_k'|=75$}. 

As a result, for AM-RELDEC the total amount of training data amounts to $|\mathscr{L}|+|\mathscr{L}_k|+|\mathscr{L}_k'|=14925$, which is less than $15000$ since the global policy is not adapted during the $100$-th meta-learning step (online learning). Note that the training set sizes $|\mathscr{L}|$ and $|\mathscr{L}|+|\mathscr{L}_k|$ are chosen to ensure that the dataset is large enough for accurate training without incurring too much computation time. Finally, we fix the maximum number of decoding iterations as $I_{\max}=50$ for the AB and 5G-NR codes, but use $I_\max=5$ for the WRAN code to enable a comparison with the hyper-network scheme of \cite{NW19}. 

For both training and inference, we transmit all-zero codewords using BPSK modulation. Note that training with the all-zero codeword is sufficient as, due to the symmetry of the BP decoder and the channel, the decoding error is independent of the transmitted signal (see e.g. \cite[Lemma 4.92]{RU08}). For performance measures, we consider both the bit error rate (BER), given by $\Pr[\hat{x}_v\neq x_v]$, $v\in[[n]]$, and the frame error rate (FER), given by $\Pr[\mathbf{\hat{x}} \neq \mathbf{x}]$. 

As benchmarks we also simulate the codes using NS \cite{CGW10}, {EDS-BP (efficient schedule layered BP) \cite{edslbp}, RBL-BP (reliability-based layered BP) \cite{rblbp}, and a VN-based layered decoding scheme proposed in the 5G 3GPP New Radio standardization process \cite{5gdoc}, respectively. For the method in \cite{5gdoc}, we choose a layer size of four for all layers, irrespective of the code (all their lengths are divisible by four). This choice ensures that the number of layers used for the  5G code is as close as possible to the number of layers for the rate $1/5$ codes used in \cite{5gdoc}.} 

%Recall that NS schedules only the CN with the highest residual per iteration, whereas EDS-LBP schedules only the VN with highest relative residual per iteration.} Hence, for fair comparison with our work, we ensure that the maximum number of NS and {EDS-LBP iterations is $m I_\max$ and $nI_\max$, respectively. One the other hand, RBL-BP schedules all VNs per iteration with LLR magnitudes smaller than some threshold, which increases by a constant $\delta$ in each iteration. We found $\delta=0.5$ (resp., $3$) for the AB and 5G-NR code (resp., WRAN code) to give the best results. The maximum number of iterations used for RBL-BP is $I_\max${, and the same is also true for VN-based layered decoding}.}

%{As a benchmark, we also take into account the state-of-the-art layered decoding scheme \cite{hocevar} where a layer refers to a distinct cluster of contiguous CNs. Each layer is decoding sequentially in every iteration, and the messages in the layer induced subgraph is updated via flooding. We consider only two layers each consisting of $m/2$ CNs.}

The BER vs. channel SNR, in terms of $E_b/N_0$ in dB, for the $(3,5)$ AB, the $[520,100]$ 5G-NR LDPC, and the $[384,256]$-WRAN code using these decoding techniques are shown in Figs. \ref{fig:res34}(a), \ref{fig:res56}(a), and \ref{fig:res12}(a), respectively, where we have limited the plots to the more interesting moderate SNR regime. {The EDS-LBP scheme outperforms NS and RBL-BP for the AB and WRAN code. However, for the 5G code, EDS-LBP is incapable of achieving a lower FER than NS since the Tanner graph of this code contains a large number of degree one VNs which have zero relative residuals. The RBL-BP decoder generally outperforms NS, but is sub-optimal compared to EDS-LBP for all codes but the 5G code, demonstrating the effectiveness of reliability-based VN scheduling for achieving a relatively low FER.} {The VN-based layered scheduling scheme of \cite{5gdoc} outperforms the RBL-BP scheme mainly due to its small layer size, which contains fewer cycles and absorbing sets, compared to the layer size of RBL-BP, which grows with the number of decoder iterations. It even outperforms EDS-LBP for the WRAN code.} 

%Note that for all codes{, the performance of D-RELDEC is similar to random scheduling mainly because the subgraphs induced by the larger clusters are more likely to contain objects detrimental to decoding, such as cycles and absorbing sets. Hence, D-RELDEC results are not shown.}  

{It is important to note that the relatively poor error-floor performances of NS and EDS-LBP are primarily due to myopic errors \cite{CGW10} caused by the presence of short cycles in the code's Tanner graph. These errors force the decoder to schedule only a subset of the nodes, \emph{i.e.,} a single node gets scheduled multiple times in the same iteration, resulting in an early error-floor. The RBL-BP decoder also schedules a subset of the VNs, leading to similar error-floor performance. In contrast random sequential decoding, layered decoding, and our RL-based decoders schedule all nodes sequentially in every iteration.} 

%the performance of D-RELDEC is worse compared to other learning schemes, 

{Finally, we can observe from Figs.~\ref{fig:res34}(a), \ref{fig:res56}(a), and \ref{fig:res12}(a)  that the RELDEC schemes clearly outperform the other decoding schemes, including another learning-base decoder, namely the hyper-network decoder of \cite{NW19} (shown only for the WRAN LDPC code). AM-RELDEC performs slightly better than RELDEC, due to the better adaptation to the channel SNR. The FER vs. SNR performance shown in Figs.~\ref{fig:res34}(b), \ref{fig:res56}(b), and \ref{fig:res12}(b) exhibits similar behavior. }
%{Additionally, this excellent decoding performance} demonstrates the agility of AM-RELDEC %for adapting local CN scheduling policies online using a very small number of training examples.}

{We also simulated the codes using a deep-Q-learning-based RELDEC scheme for $z=2$. Deep Q-learning was necessary in this case, as the size of the Q-table grows exponentially in the number of cluster neighbors $l_a$ (which increases with $z$) making it infeasible to learn through  an explicit Q-table. We also employed AM-RELDEC by using  online training with only $7$ training samples instead of $75$. The former performs similarly to random sequential decoding, whereas the latter performs only slightly worse than AM-RELDEC with $75$ training samples, but better than RELDEC. Thus, their performances are not shown in the figures.}
In Table \ref{tab:tab}, we compare the average{\footnote{The average is computed over the total number of channel simulations for a fixed SNR.}} number of CN to VN messages propagated in the considered decoding schemes to achieve the results in Figs.~\ref{fig:res34}-\ref{fig:res12}. {Note that in EDS-LBP, RBL-BP, {and the VN-based layered decoder of \cite{5gdoc},} the subgraph  induced by the scheduled VN includes all its neighboring CNs, and all the neighboring VNs of those CNs. When a VN is scheduled, the neighboring CNs of the scheduled node propagate messages to the other VNs in the subgraph. As a result, when $n$ distinct VNs are scheduled in EDS-LBP or RBL-BP (which is equivalent to one iteration of flooding), a VN that is common to the subgraphs of the scheduled nodes is updated more than once in the same iteration. Consequently, EDS-LBP, RBL-BP{, and the layered decoder of \cite{5gdoc}} can be more complex than flooding.} 
We also observe  that both RELDEC and AM-RELDEC, on average, generate a lower number of CN to VN messages when compared to the other decoding schemes, offering a significant reduction in message-passing complexity for moderate-length LDPC codes. Moreover, RELDEC-based schemes do not require the complex residual calculation of NS and EDS-LBP.

%\vspace{-0.5cm}
\begin{figure}[h]
\begin{subfigure}{.5\textwidth}
  \centering
  %\centerline{\resizebox{2.5in}{2in}{\includegraphics{pics/res1.pdf}}}
  \includegraphics[scale=0.46]{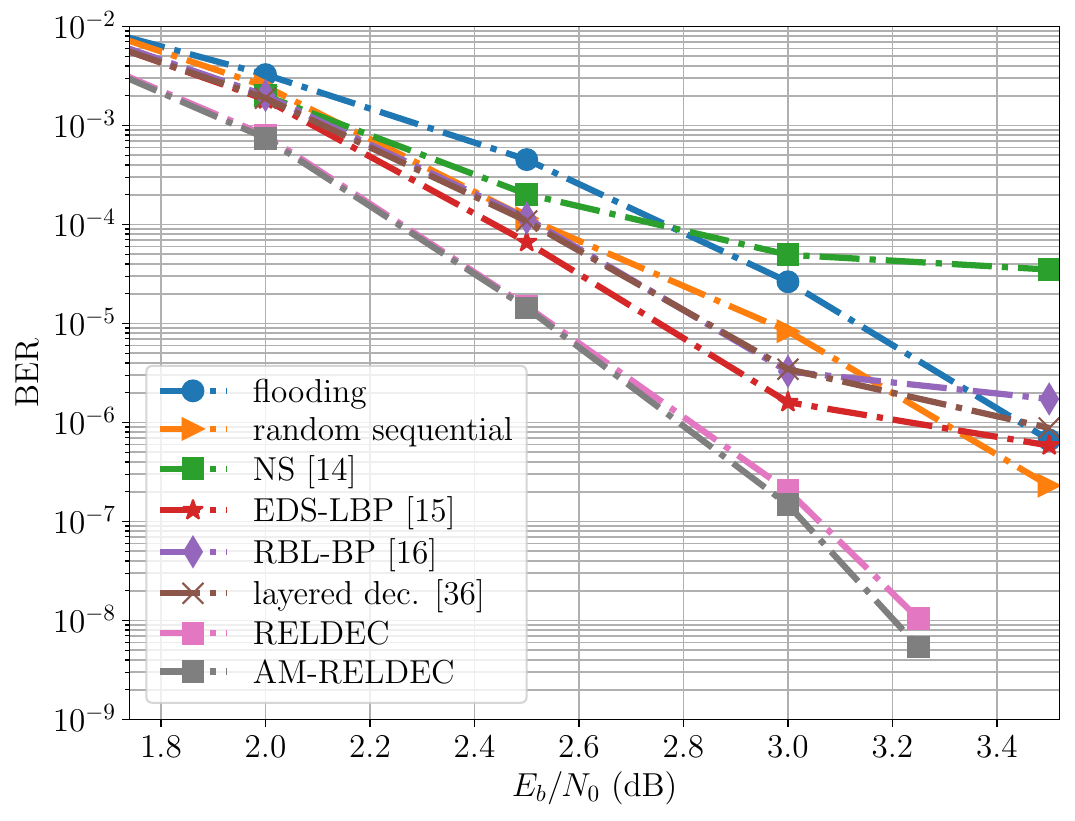}
  \caption{}
  %\label{fig:res3}
\end{subfigure}
\hfill
\begin{subfigure}{.5\textwidth}
  \centering
  %\centerline{\resizebox{2.5in}{2in}{\includegraphics{pics/res2.pdf}}}
  \includegraphics[scale=0.46]{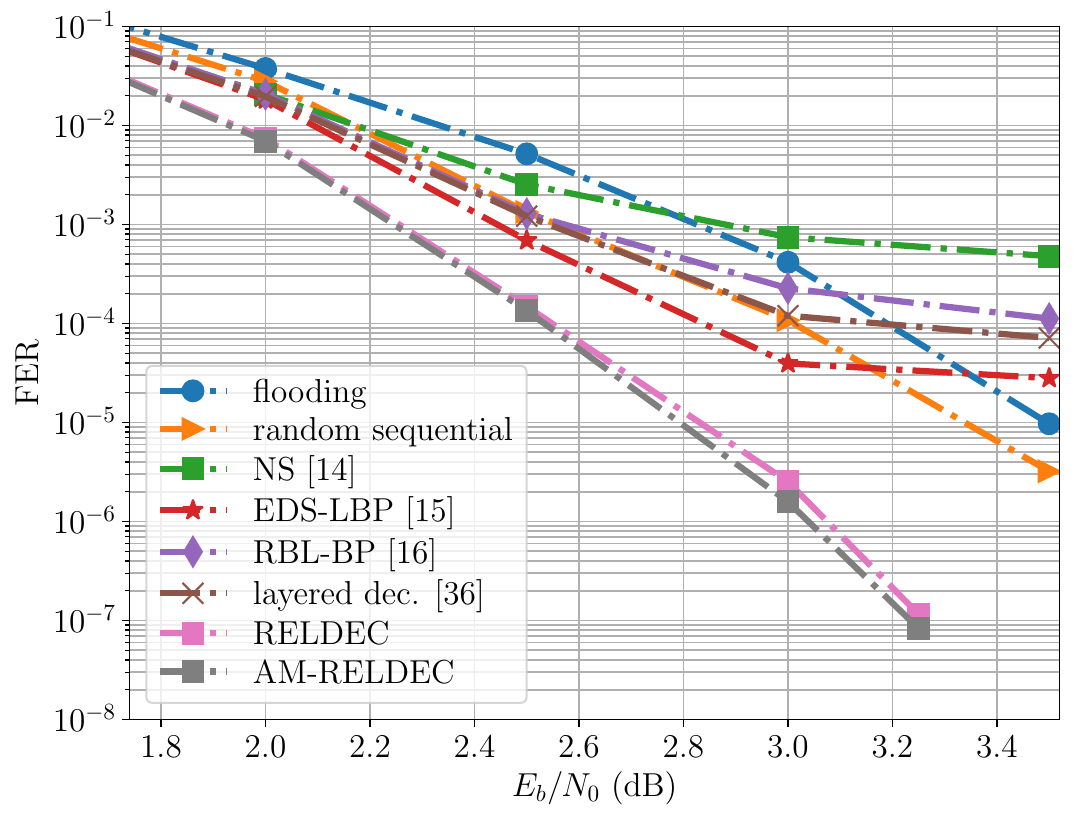}
  \caption{}
  %\label{fig:res4}
\end{subfigure}
\vspace{-2ex}
\caption{BER and FER results using different BP decoding schemes for a $(3,5)$ AB-LDPC code with block length $n=500$ and $I_{\max}=50$ decoder iterations for an AWGN channel.}
\vspace{1ex}
\label{fig:res34}
\end{figure}

%\vspace{-0.5cm}
\begin{figure}[h]
\begin{subfigure}{.5\textwidth}
  \centering
  %\centerline{\resizebox{2.5in}{2in}{\includegraphics{pics/res1.pdf}}}
  \includegraphics[scale=0.475]{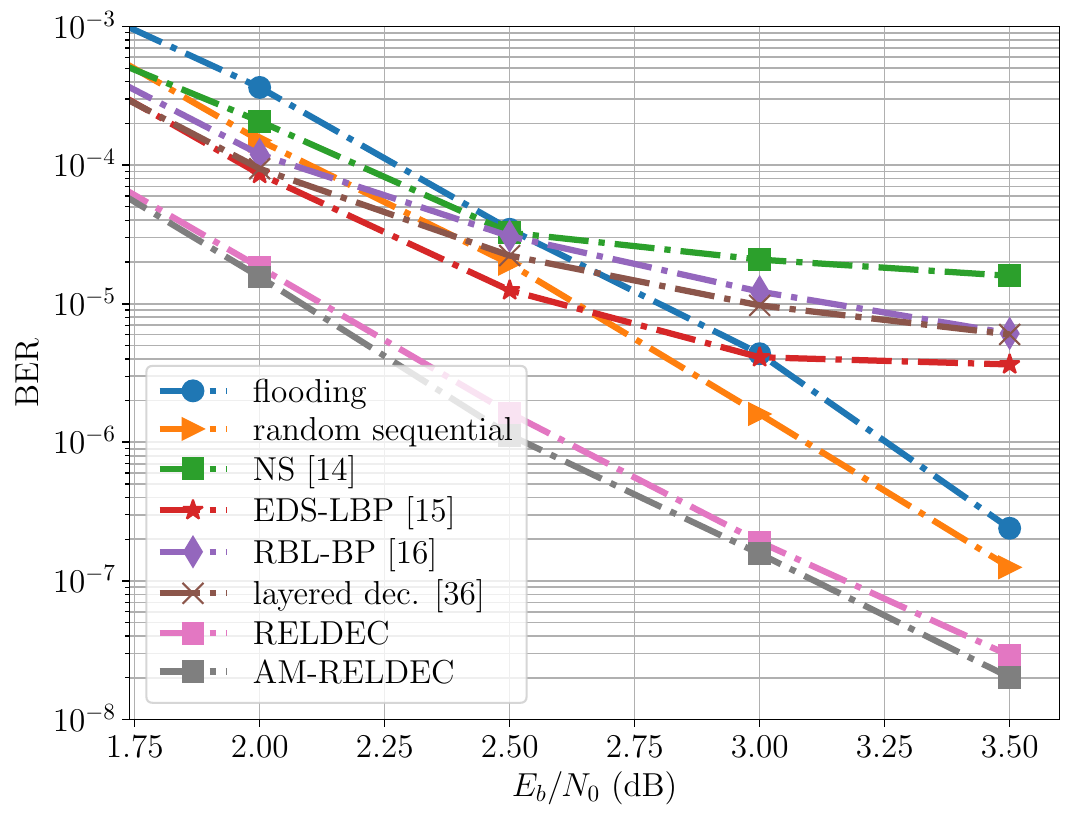}
  \caption{}
  %\label{fig:res5}
\end{subfigure}
\hfill
\begin{subfigure}{.5\textwidth}
  \centering
  %\centerline{\resizebox{2.5in}{2in}{\includegraphics{pics/res2.pdf}}}
  \includegraphics[scale=0.475]{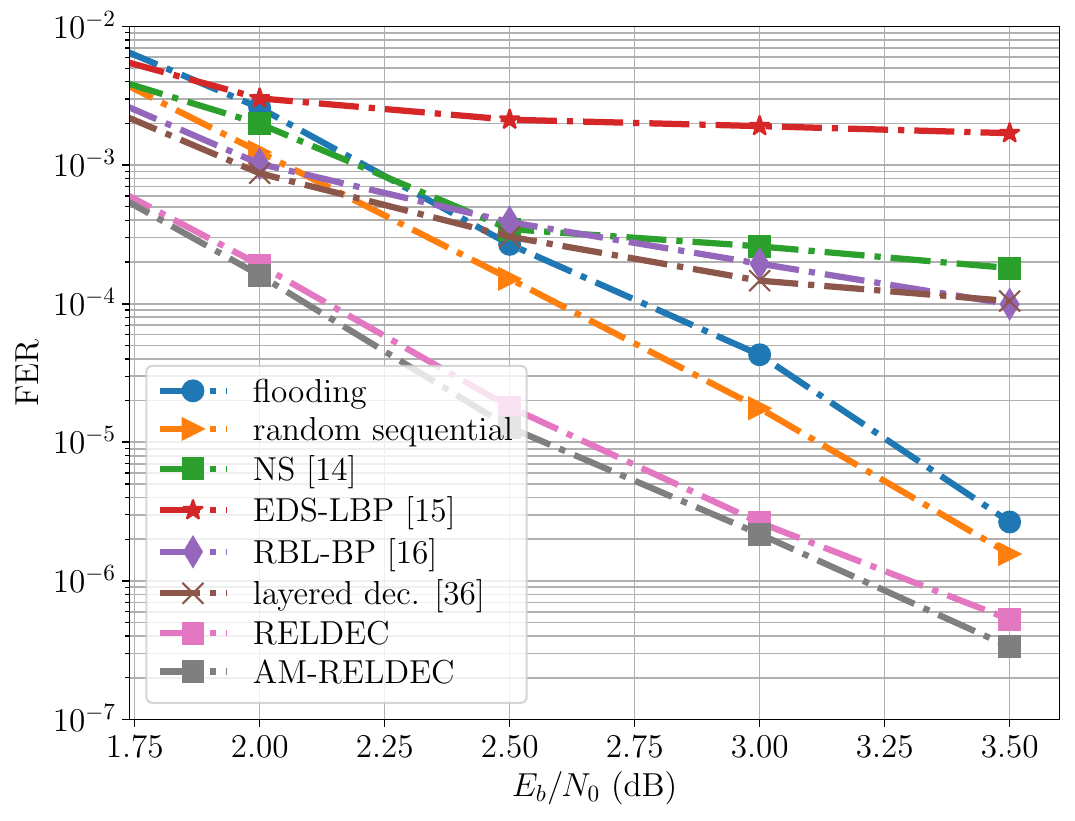}
  \caption{}
  %\label{fig:res6}
\end{subfigure}
\vspace{-2ex}
\caption{BER and FER results using different BP decoding schemes for a $[520,100]$ 5G-NR LDPC code and $I_{\max}=50$ decoder iterations for an AWGN channel.}
\vspace{1ex}
\label{fig:res56}
\end{figure}

%\vspace{-0.5cm}
\begin{figure}[h]
\begin{subfigure}{.5\textwidth}
  \centering
  %\centerline{\resizebox{2.5in}{2in}{\includegraphics{pics/res1.pdf}}}
  \includegraphics[scale=0.475]{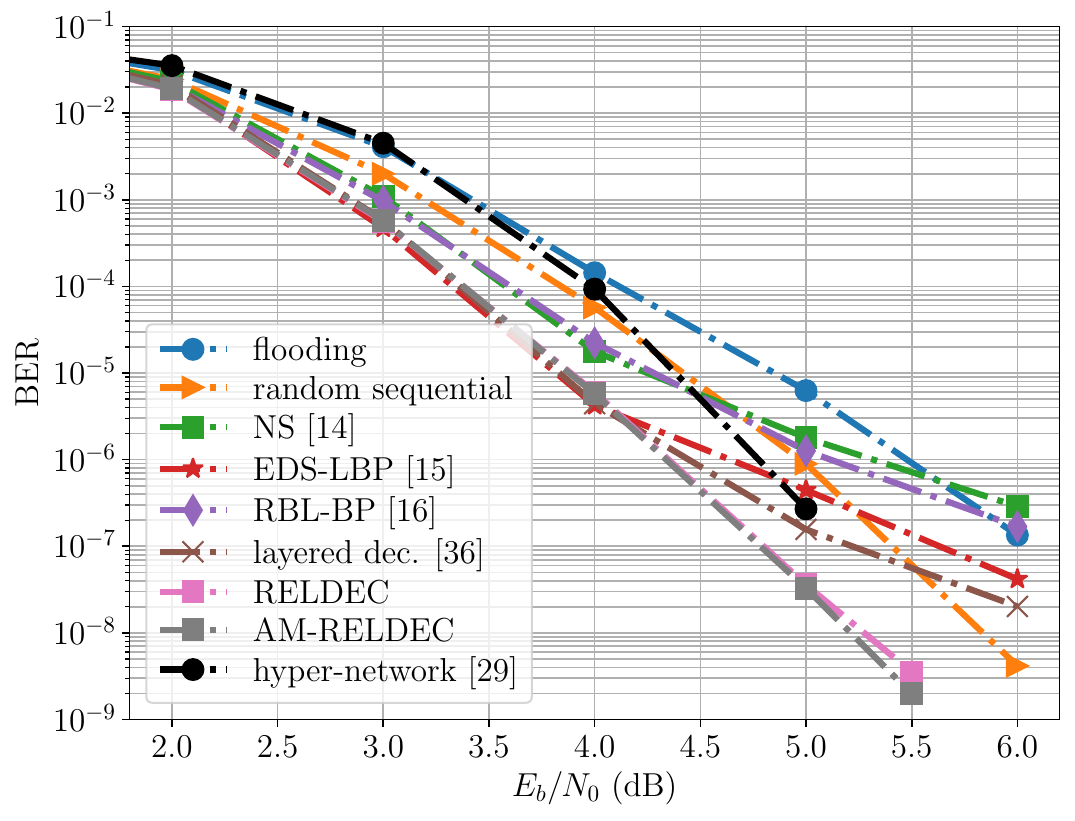}
  \caption{}
  %\label{fig:res1}
\end{subfigure}
\hfill
\begin{subfigure}{.5\textwidth}
  \centering
  %\centerline{\resizebox{2.5in}{2in}{\includegraphics{pics/res2.pdf}}}
  \includegraphics[scale=0.475]{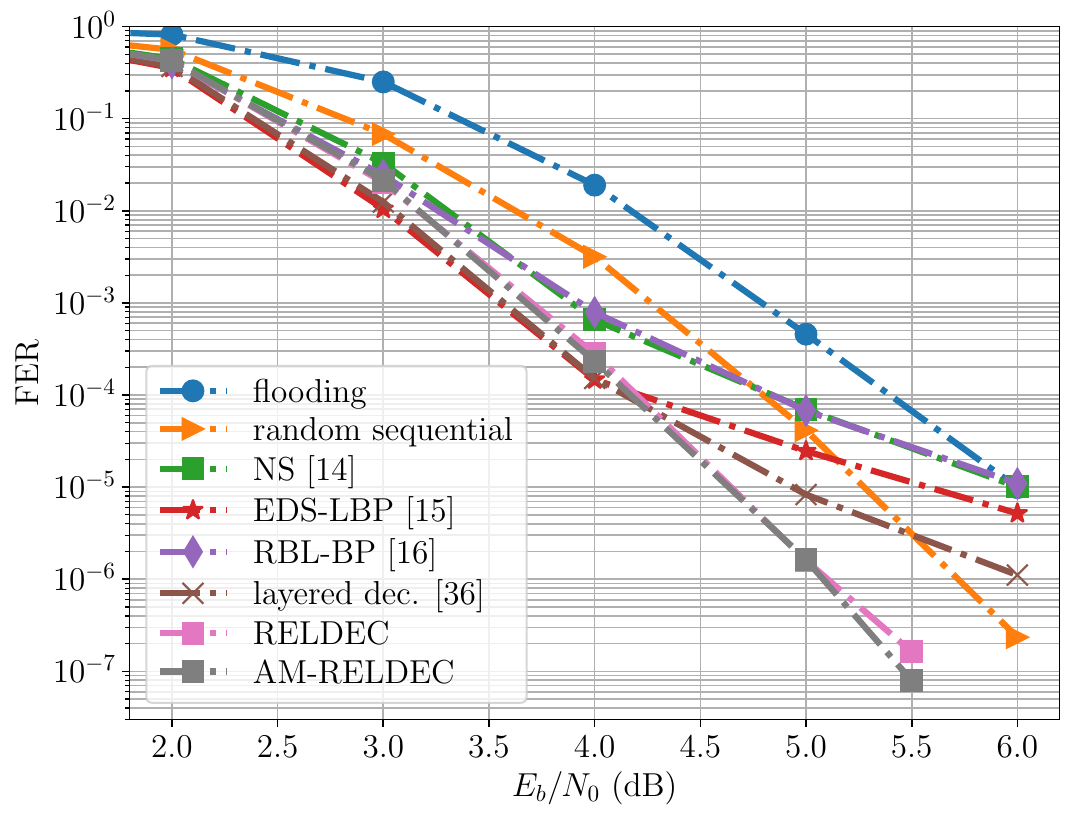}
  \caption{}
  %\label{fig:res2}
\end{subfigure}
\vspace{-2ex}
\caption{BER and FER results using different BP decoding schemes for a $[384,256]$-WRAN LDPC code and $I_{\max}=5$ decoder iterations for an AWGN channel. The BER results for the hyper-network scheme are taken from \cite{NW19}.}
\vspace{1ex}
\label{fig:res12}
\end{figure}

%\vspace{-0.5cm}
{\linespread{1}\selectfont 
\begin{table*}[h]
\tiny
%\vspace{-1ex}
\centering
\resizebox{\textwidth}{!}{
%\resizebox{\columnwidth}{!}{
	\begin{tabular}{|c|c|c|c|}
		\hline
		\textbf{SNR (dB)}
		&$\mathbf{2}$
		&$\mathbf{2.5}$ 
		&$\mathbf{3}$ \\
		
		%\hline=\\
		%\cline{1-7}
		
		\hline\hline
		flooding&$16409$ $(12752)$&$10742$ $(10745)$&$8123$ $(9491)$\\
		\hline
		random scheduling &$10533$ $(7580)$&$6436$ $(6598)$&$4850$ $(5977)$\\
		\hline
		NS &$7628$ $(7671)$&$5404$ $(7322)$&$4887$ $(7095)$\\
		\hline
		{EDS-LBP} &{$14242$ $(17250)$}&{$4948$ $(14144)$}&{$3281$ $(12664)$}\\
		\hline
		{RBL-BP} &{$28556$ $(30904)$}&{$18813$ $(30592)$}&{$15312$ $(31346)$}\\
		\hline
		%{D-RELDEC} &{$10750$ $(8044)$}&{$6571$ $(6724)$}&{$4965$ $(6089)$}\\
		{VN-based layered dec.} &{$19108$ $(17706)$}&{$9580$ $(16539)$}&{$7143$ $(15505)$}\\
		\hline
		RELDEC &$6601$ $(5771)$&$4821$ $(5131)$&$4028$ $(4619)$\\
		\hline
		{AM-RELDEC} &$6440$ $(5763)$&$4725$ $(5128)$&$3956$ $(4615)$\\
		\hline
	\end{tabular}
	\quad
	\begin{tabular}{|c|c|c|c|}
		\hline
		\textbf{SNR (dB)}
		&$\mathbf{3}$
		&$\mathbf{4}$ 
		&$\mathbf{5}$ \\
		
		%\hline=\\
		%\cline{1-7}
		
		\hline\hline
		flooding&$5171$&$3355$&$2184$\\
		\hline
%		random ($z=3$)&$44338$&$11102$&$5005$\\
%		\hline
%		RL ($z=3$)&$40448$&$10694$&$4998$\\
%		\hline
%		random ($z=2$)&$36328$&$10254$&$4994$\\
%		\hline
%		RL ($z=2$)&$31383$&$7349$&$4225$\\
		random scheduling &$3506$&$2207$&$1546$\\
		\hline
		{NS} &$3054$&$2625$&$2483$\\ 
		\hline
		{EDS-LBP} &{$4164$}&{$1745$}&{$1119$}\\
		\hline
		{RBL-BP} &{$12218$}&{$7367$}&{$4723$}\\
		\hline
		%{D-RELDEC} &{$3578$}&{$2236$}&{$1575$}\\
		{VN-based layered dec.} &{$12918$}&{$11862$}&{$11840$}\\
		\hline
		RELDEC &$3193$&$2206$&$1584$\\
		\hline
		{AM-RELDEC} &$3203$&$2206$&$1585$\\
		\hline
	\end{tabular}
}
	\caption{Average number of CN to VN messages propagated in various decoding schemes for the  $(3,5)$ AB ($[520,420]$ 5G-NR) LDPC code, shown on the left, and the $[384,256]$-WRAN LDPC code, shown on the right, to attain the results shown in Figs.~\ref{fig:res34}-\ref{fig:res12}. {The results are obtained via Monte Carlo simulations based on at least $300$ frame errors over all shown SNR values.}}
	\label{tab:tab} 
\end{table*}
}

\vspace{-1ex}
\section{Conclusion}
\vspace{-0.5ex}
\label{sec:concl}
%{\linespread{0.97}\selectfont
We presented RELDEC, a novel RL-based sequential decoding scheme proposed to optimize the scheduling of CN clusters for moderate length LDPC codes. In contrast to our previous work, the main contributions of this work include a new complexity-reduced state space model built using the collection of possible outputs of individual clusters {as well as} a scheduling approach that updates all CN clusters sequentially within each decoder iteration. Furthermore, we propose novel meta-learning based sequential decoding schemes, {namely AM-RELDEC,} to further improve the decoding performance with respect to RELDEC. {The learning flexibility of AM-RELDEC allows the decoder to quickly adapt to changing channel conditions due to fading, making it well-suited, with moderate computational and memory complexity, to error correction in wireless communications scenarios.} Experimental results show that by learning the cluster scheduling order using RELDEC and its meta-learning counterparts, we can significantly outperform flooding and random scheduling schemes without any expensive computation of residuals as in {previous} sequential scheduling schemes. Our presented performance gains include lowering both BER and message-passing complexities.  {We believe that the observed gains for the codes under consideration will carry over to longer codes, provided the additional training complexity can be realized.}

\vspace{-4ex}

%\newpage
%\mbox{~}
\bibliographystyle{IEEEtran}
%{\linespread{0.9}\selectfont 
\bibliography{Bib_journal_latest}
%}

%\vspace{-0.5cm}
%\begin{figure}[h]
%  \centering
%  %\centerline{\resizebox{3in}{2.4in}{\includegraphics{pics/res1d.pdf}}}
%  \includegraphics[scale=0.4]{pics/res1.pdf}
%  \vspace{-0.12cm}
%  \caption{BER results using different BP decoding schemes for a {$[384,256]$-WRAN} LDPC code with block length {$n=384$}.} %compared with the performance of a %BCH($128,64$) code.}
%  \label{fig:res1}
%\end{figure}
%
%%\vspace{-0.5cm}
%\begin{figure}[h]
%  \centering
%  %\centerline{\resizebox{3in}{2.4in}{\includegraphics{pics/res2d.pdf}}}
%  \includegraphics[scale=0.4]{pics/res2.pdf}
%  \vspace{-0.12cm}
%  \caption{FER results using different BP decoding schemes for a {$[384,256]$-WRAN} LDPC code with block length {$n=384$}.}
%  %compared with the performance of a %BCH($128,71$) code.}
%  \label{fig:res2}
%\end{figure}

%\begin{figure}[h]
%  \centering
%  %\centerline{\resizebox{3in}{2.4in}{\includegraphics{pics/res1d.pdf}}}
%  \includegraphics[scale=0.4]{pics/res3.pdf}
%  \vspace{-0.12cm}
%  \caption{BER results using different BP decoding schemes for a $(3,5)$ AB-LDPC code with block length $n=500$.} %compared with the performance of a %BCH($128,64$) code.}
%  \label{fig:res3}
%\end{figure}
%
%%\vspace{-0.5cm}
%\begin{figure}[h]
%  \centering
%  %\centerline{\resizebox{3in}{2.4in}{\includegraphics{pics/res2d.pdf}}}
%  \includegraphics[scale=0.4]{pics/res4.pdf}
%  \vspace{-0.12cm}
%  \caption{FER results using different BP decoding schemes for a $(3,5)$ AB-LDPC code with block length $n=500$.}
%  %compared with the performance of a %BCH($128,71$) code.}
%  \label{fig:res4}
%\end{figure}

\end{document}